\documentclass[fleqn,usenatbib]{mnras}

\usepackage{newtxtext,newtxmath}
\usepackage{soul}

\usepackage[T1]{fontenc}

\DeclareRobustCommand{\VAN}[3]{#2}
\let\VANthebibliography\thebibliography
\def\thebibliography{\DeclareRobustCommand{\VAN}[3]{##3}\VANthebibliography}

\usepackage{threeparttable}

\usepackage{graphicx}	
\usepackage{amsmath}	
\usepackage[normalem]{ulem}

\usepackage{tikz,xcolor,hyperref}

\definecolor{lime}{HTML}{A6CE39}
\DeclareRobustCommand{\orcidicon}{
\begin{tikzpicture}
\draw[lime, fill=lime] (0,0)
circle[radius=0.13]
node[white]{{\fontfamily{qag}\selectfont \tiny \.{I}D}}; 

\end{tikzpicture}

\hspace{-2mm}

}
\foreach \x in {A, ..., Z}{%
\expandafter\xdef\csname orcid\x\endcsname{\noexpand\href{https://orcid.org/\csname orcidauthor\x\endcsname}{\noexpand\orcidicon}}
}

\usepackage{url}
\hypersetup{hidelinks,
	colorlinks=true,
	allcolors=blue,
	pdfstartview=Fit,
	breaklinks=true}
	
\title[The 850 $\micron$ number counts in the SSA22 field]{Deep Submillimetre and Radio Observations in the SSA22 Field. II. Sub-millimetre source catalogue and number counts}

\author[Xin Zeng, Yiping Ao and Yuheng Zhang]
{
Xin Zeng\hspace{-1.mm}\orcidA{},$^{1,2}$
Yiping Ao\hspace{-1.mm}\orcidB{}$^{1,2}$\thanks{E-mail:ypao@pmo.ac.cn}
and Yuheng Zhang\hspace{-1.mm}\orcidC{}$^{1,2,3,4}$
\\
$^{1}$Purple Mountain Observatory, Chinese Academy of Sciences, 10 Yuanhua Road, Nanjing, Jiangsu 210023, People's Republic of China\\
$^{2}$School of Astronomy and Space Sciences, University of Science and Technology of China, Hefei, Anhui 230026, People's Republic of China\\
$^{3}$Instituto de Astrofísica de Canarias (IAC), E-38205 La Laguna, Tenerife, Spain\\
$^{4}$Universidad de La Laguna, Dpto. Astrofísica, E-38206 La Laguna, Tenerife, Spain
}

\date{Accepted 2023 December 14. Received 2023 December 8; in original form 2023 August 28}

\pubyear{2023}

\begin{document}
\label{firstpage}
\pagerange{\pageref{firstpage}--\pageref{lastpage}}
\maketitle

\begin{abstract}
We present the deepest 850 $\upmu$m map of the SSA22 field to date, utilizing a combination of new and archival observations taken with SCUBA-2, mounted at the James Clerk Maxwell Telescope (JCMT). The mapped area covers an effective region of approximately 0.34 deg$^2$, achieving a boundary sensitivity of 2 mJy beam$^{-1}$, with the deepest central coverage reaching a depth of $\sigma_\text{rms}$ $\sim$ 0.79 mJy beam$^{-1}$, the confusion noise is estimated to be $\sim$ 0.43 mJy beam$^{-1}$.
A catalogue of 850 $\upmu$m sources in the SSA22 field is generated, identifying 390 sources with single-to-noise ratios above 3.5, out of which 92 sources exceed 5$\sigma$. 
The derived intrinsic number counts at 850 $\upmu$m are found to be in excellent agreement with published surveys. 
Interestingly, the SSA22 number counts  also exhibit an upturn in the brighter flux region, likely attributed to local emitters or lensing objects within the field. 
On the scale of $\sim$ 0.3 deg$^2$, the 850 $\upmu$m number counts are unaffected by cosmic variance and align with the blank field.
In the deep region ($\sigma_\text{rms}$ $\leqslant$ 1 mJy), the counts for fluxes below 8 mJy are consistent with the blank field, and the excess in the brighter regime is not significant.
Due to the limited number of very bright sources and the insubstantial cosmic variance in our field, we attribute the fluctuations in the number counts primarily to Poisson noise.
The SCUBA-2 850 $\upmu$m detection in the SSA22 field does not exhibit indications of overdensity.

\end{abstract}

\begin{keywords}
catalogues -- galaxies: high-redshift -- galaxies: star formation -- submillimetre: galaxies
\end{keywords}



\section{Introduction}

The exploration of the formation of the first stars/galaxies and the evolution of galaxies is crucial for our understanding of the universe. In the 1980s, the \textit{Infrared Astronomical Satellite} (\textit{IRAS}) made a significant discovery, revealing that the bolometric luminosities of some galaxies were predominantly dominated by far-infrared emission \citep{sanders_mirabel1996, simpson2019}. Subsequently, the space-based \textit{Cosmic Background Explorer} (\textit{COBE}), aimed at studying the cosmic background radiation, found a substantial portion of the background radiation was emitted in the far-infrared and submillimetre regime \citep{puget1996,fixsen1998,hauser1998,casey2014,zavala2017}.
These observations of galaxies suggested that the interstellar dust absorbed ultraviolet radiation produced during intense star formation and reradiated it at far-infrared wavelengths \citep{hildebrand1983,dole2006,simpson2019}. Consequently, due to spectral redshift \citep{garratt2023}, this radiation was detected in the submillimetre band, leading to the identification of a population of dusty star-forming galaxies known as submillimetre galaxies (SMGs).

Despite severe water vapour absorption in the far-infrared band, some submillimetre atmospheric windows remain survived, enabling the observation of SMGs. Additionally, submillimetre observations benefit from a strong negative K correction, resulting in luminosities of submillimetre galaxies remaining nearly constant regardless of their redshift. Specifically, for the 850 $\upmu$m detection, the radiation from thermal dust approaches the Rayleigh–Jeans limit, and as redshift increases, the rest-frame band shifts toward the peak of the spectral energy distribution (SED) of galaxy, compensating for cosmological dimming \citep{blain2002,dale_helou2002,casey2014,geach2017,garratt2023}. For instance, submillimetre galaxies exhibit similar flux densities whether they are located at a redshift of 0.5 or 10 \citep{blain1993}.
Therefore, the 850 $\upmu$m observation offers a significant advantage for tracing back to the epoch of cosmic reionization.

In the late 1990s, the Submillimetre Common User Bolometer Array (SCUBA) mounted on the 15 m James Clerk Maxwell Telescope (JCMT) at Mauna Kea captured the first glimpses of submillimetre galaxies \citep{smail1997,barger1998,hughes1998,lilly1999}. Subsequently, the new generation of the bolometer array camera, SCUBA-2, advanced our understanding of the properties and physical processes of SMGs.
The redshift distribution of SMGs selected at 850 $\upmu$m exhibit a peak $\sim$ 2 - 3 \citep{chapman2005,wardlow2011,chen2013a,zavala2017}. It is essential to note that the redshift distribution of SMGs depends on the observation wavelength, with galaxies selected in longer wavebands tending to have higher redshifts \citep{chapin2009,chen2013a,roseboom2013,zavala2014,wang2017}.
Regarding their properties, these galaxies are heavily dusty and have masses on the order of $\sim$ 10$^{10-11}$ M$_\odot$ \citep{swinbank2004, hainline2011, michalowski2012}, and contain large amounts of gas \citep{frayer1998, greve2005, ivison2011, thomson2012}. SMGs are luminous in the infrared, with luminosities on the order of $\sim$ 10$^{12-13}$ L$_\odot$ \citep{barger1998, zavala2017, hyun2023}, and their star formation rates usually range between $\sim$ 100 - 1000 M$_\odot$ yr$^{-1}$ \citep{magnelli2012, swinbank2014, Barger2014}.
Despite these observations, the exact formation mechanisms of SMGs remain unclear. Leading hypotheses include starbursts triggered by major merger phases of gas discs in galaxies \citep{baugh2005, ivison2012} and galaxy interactions \citep{swinbank2010, kartaltepe2012, chen2015, wang2017}. Cosmological simulations have also inspired other formation mechanisms, such as long-period gas accretion \citep{narayanan2015} and disc instabilities \citep{lacey2016}.

X-ray detections of submillimetre galaxies selected at 850 $\upmu$m revealed that approximately 20 per cent of these galaxies host active galactic nuclei (AGNs) \citep{alexander2005, almaini2005, pope2008, wang2017}, some SMGs had been found to reside in quasars, which are further associated with supermassive black holes (SMBHs) \citep{jones2017,alexander2008,wang2013}.
Due to their nature and characteristics, SMGs were often considered as progenitors of massive elliptical galaxies in the local universe \citep{eales1999, lilly1999, genzel2003, simpson2017}. Observations had suggested that these galaxies were located in overdense regions \citep{dressler1980, menendez-delmestre2013, dannerbauer2014, battaia2018}. While they appeared to trace the large-scale structure in the high-z universe, their individual connection to clusters or protoclusters remained unclear \citep{umehata2014}.

The SSA22 field, originally one of the four target fields of the deep imaging and spectroscopic survey, derived its name from `Small Selected Area (SSA) at 22$^\text{h}$' \citep{lilly1991, cowie1994}. \citet{steidel1998} reported a highly significant concentration of galaxies at a redshift of z $\sim$ 3.09 in the SSA22 field, traced by Lyman-break galaxies (LBGs). Subsequent studies of Lyman $\alpha$ emitters (LAEs) supported the existence of an overdensity region in this field, with the local density being 6 times the average, indicating a robust `protocluster' \citep{steidel2000, hayashino2004, tamura2009, yamada2012, umehata2014}. \citet{umehata2019} also discovered a large-scale filamentary structure in the core of the SSA22 protocluster.
Several studies supported the synchronous formation of SMGs and LAEs within the same cosmic structure, implying that SMGs might be also preferentially formed in dense regions \citep{tamura2009, umehata2015}, where star formation is fuelled and black holes are grown \citep{lehmer2009, umehata2019}.
A population of extended LAEs was referred to as Lyman $\alpha$ blobs (LABs), which might preferentially reside in dense environments at high redshift, hinting at an association with massive galaxy formation \citep{matsuda2004, ao2017}. 
The origin of LABs remained unclear  and might be diverse, but some observations suggested that they were associated with large-scale gas outflows driven by AGN or intense starbursts, and filamentary LABs might be linked to cold accretion streams from the surrounding intergalactic medium (IGM) \citep{matsuda2011, geach2014, ao2015, ao2020}.

The paper is organized as follows. In Section~\ref{sec:observation_reduction}, we provide a detailed description of our observations and data reduction process in the SSA22 field. Section~\ref{sec:map_extraction} presents the SSA22 coverage map and outlines the source extraction procedure. We elaborate on the Monte Carlo simulation approach used to correct the extracted information of the submillimetre sources and present our source catalogue in Section~\ref{sec:simulation}. Finally, we report the number counts for the SSA22 field and discuss the results in Section~\ref{sec:counts_and_discussion}. The article employs the cosmological parameters: H$_0$ = 67.8 km$^{-1}$ s$^{-1}$ Mpc$^{-1}$, $\Omega_{\Lambda}$ = 0.692, $\Omega_m$ = 0.308.

\section{OBSERVATIONS AND DATA REDUCTION}
\label{sec:observation_reduction}

\subsection{Observations}
A subregion of the SSA22 field was observed with SCUBA-2 on the JCMT between 2015 April and June \citep[program ID: M15AI91;][]{ao2017}. 
The submillimetre transmission quality heavily relied on the atmospheric extinction, quantified by the precipitable water vapour (PWV), which was obtained from a 183 GHz water vapour monitor (WVM) mounted on JCMT \citep{holland2013}. The PWV was directly converted to the zenith opacity at 225 GHz ($\tau_{225}$), a crucial factor for extinction correction of the observation frequency \citep{dempsey2013, mairs2021}. 
The weather conditions during the observation ranged from 0.04 to 0.08, with an integral time of 22 per cent categorized as Grade 1 for $\tau_\text{225}$ < 0.05 (PWV < 0.83 mm), and the remaining categorized as Grade 2 for 0.05 < $\tau_\text{225}$ < 0.08 (0.83 mm < PWV < 1.58 mm). 
The quality of our data was also influenced by the observing elevation. To ensure an adequate exposure time for the map, an elevation restriction of under 70$\degr$ \footnote{\url{https://www.eaobservatory.org/jcmt/instrumentation/continuum/scuba-2/observing-modes/##High_elevation_constraints}} was adopted. The elevation of our observations ranged from 44$\degr$ to 69$\degr$. 

The new observation map was centred at $\alpha$ = 22$^\text{h}$17$^\text{m}$31$\fs$70, $\delta$ = +00$\degr$17$\arcmin$50$\farcs$0 (J2000), with a total on-source mapping time of 19 h.
To achieve a larger coverage area with a smoothly increasing noise gradient, we employed a rotating PONG pattern. In this observation mode, the telescope scanned across a predetermined sky region and `bounced' off the edge. At the end of each `bounce', the map was rotated, and the PONG pattern repeated the motion \citep{geach2017}. 
This scanning method was available in various sizes (e.g., 900$\arcsec$, 1800$\arcsec$, 2700$\arcsec$, 3600$\arcsec$, and 7200$\arcsec$) to meet different coverage requirements. For our observations, we used the PONG-900 observing strategy with a scan velocity of 280$\arcsec$ s$^{-1}$, which repeated 11 times during an approximate 40-minute integration time\footnote{\url{https://www.eaobservatory.org/jcmt/instrumentation/continuum/scuba-2/observing-modes/}}, covering a field of diameter 15 arcmin.

The SSA22 field had been previously mapped as one of the target fields of the SCUBA-2 Cosmology Legacy Survey \citep[S2CLS;][]{geach2017}. The earlier observation (program ID: MJLSC02) adopted the PONG-1800 scanning mode at the centre of the SSA22 field, coordinated at $\alpha$ = 22$^\text{h}$17$^\text{m}$36$\fs$30, $\delta$ = +00$\degr$19$\arcmin$22$\farcs$7 (J2000). It took 72 hours between 2012 September and 2013 December, with a sky opacity ($\tau_\text{225}$) ranging from 0.05 to 0.114.
Our new observation map achieved an average 1$\sigma$ depth of 1.1 mJy beam$^{-1}$, which is comparable to the SSA22 archive map. Combining our new observations with the archival data from the Canadian Astronomy Data Centre (CADC), with a total exposure time of 91 h, allowed us to generate the deepest submillimetre 850 $\upmu$m map of the SSA22 field to date.

The SCUBA-2 records information simultaneously from two wavebands: 450 $\upmu$m and 850 $\upmu$m. However, the 450 $\upmu$m data are less stable for analysis compared to the 850 $\upmu$m data due to its higher sensitivity to transmission fluctuations. 
Observations at 450 $\upmu$m requires much drier conditions as the atmosphere is more opaque at this wavelength, and the background radiation in the sky is more severe than at 850 $\upmu$m. Furthermore, the telescope surface is not optimized for 450 $\upmu$m \citep{coppin2006}, leading to a beam pattern that is not well described by a two-dimensional Gaussian and is very sensitive to even small JCMT dish deformations \citep{knudsen2008}. 
As a result, the data quality of 450 $\upmu$m is generally inferior to that of 850 $\upmu$m, and 450 $\upmu$m data are typically used as a supplement to the 850 $\upmu$m data. For this study, we solely present the 850 $\upmu$m data, catalogue, and results of the SSA22 field.

\subsection{Data reduction}
\label{sec:data_reduction}
During the observation, SCUBA-2 records information from the astrophysical signal, atmosphere, ambient emission, and noise in the 5120 bolometers distributed across four sub-arrays. The raw scan data is a time-varying signal stream typically split into many sub-scans of about 30 seconds, which are saved in the $\scriptsize\text{STARLINK}$ NDF format with a `.sdf' extension \citep{holland2013}.

The SCUBA-2 850 $\upmu$m data were reduced using the Dynamical Iterative Map Maker ($\scriptsize\text{DIMM}$) of the Sub-millimeter User Reduction
Facility ($\scriptsize\text{SMURF}$) \citep{chapin2013} tool package in the $\scriptsize\text{STARLINK}$ 2021A.
We specifically used the `$\textit{makemap}$' command and the configuration file `$\textit{dimmconfig\_blank\_field.lis}$' that was designed to detect extremely low significant sources, and we set the pixel size to 1 arcsec.
The map-maker started with pre-processing, independently cleaning the raw data for each observation. All segments of individual scans, approximately 40 minutes each, were connected to create continuous time-series. The time-streams were then subjected to a flat-field correction \citep{dempsey2013} that bracketed each observation and calibrated the data in units of pW using the associated fast-flat scan, and were re-sampled to match the desired map pixel size of 1 arcsec.
Next, the spikes in the data were identified and removed if the bolometer values were higher than 10 times the local noise level within a box size of 50 timeslices \citep{chapin2013}. These spikes had high amplitude and short duration in each bolometer. Sudden steps in the baseline were corrected, and any gaps were filled.
Finally, the baseline of each time-stream was subtracted using polynomial fitting. To eliminate the low-frequency noise, a high-pass filter was applied, corresponding to 200 arcsec at 850 $\upmu$m. This step was taken to avoid convergence problems caused by the high-pass filter during the iteration stage \citep{thomas_scuba-2}.

When the cleaning of time-stream data was completed, the data reduction entered the iteration stage. Firstly, the common-mode signal, primarily dominated by variations in atmospheric emission and temperature, was removed by averaging values of working bolometers of the sub-array at each time-step \citep{chapin2013}. 
Secondly, an extinction correction was applied based on the opacity monitored by the WVM. Next, the time-series was re-gridded to celestial coordinate, and the astronomical signal mode was produced and subtracted from the time-series data, which was then reprojected back to the time domain. 
In the final step, the noise mode for each bolometer was measured from the data with the removed signal mode, and this noise was used to weight each map in the mosaic step \citep{chapin2013, geach2017, thomas_scuba-2}. 
The iterative program continued looping until the maximum iteration number of 4 was reached.

After obtaining a set of individual reduced maps from the map-maker, several additional post-reduction steps needed to be performed. 
We used the Pipeline for Combining and Analysing Reduced Data ($\scriptsize\text{PICARD}$) of $\scriptsize\text{STARLINK}$ and the recipe CALIBRATE\_SCUBA2\_DATA to calibrate the data from pW to mJy beam$^{-1}$. Since all observations were completed before 2016 November 19, a new flux calibration factor (FCF) of 525000 mJy beam$^{-1}$ pW$^{-1}$ was applied simultaneously to the flux and rms layer of each map, as recommended in the new version 2021A of $\scriptsize\text{STARLINK}$ \citep{thomas_scuba-2}. This FCF had an uncertainty of 7 per cent \citep{mairs2021}. The calibration method of beam FCF, also known as peak FCF, integrated the total flux density of a source distributed over the beam region into one pixel.
We noticed that a single scan was skipped by the program because it only had 25 bolometers, and it was ignored in the remainder of the processes. Alternatively, if we used the advanced science pipeline $\scriptsize\text{ORAC-DR}$ and the recipe REDUCE\_SCAN\_FAINT\_POINT\_SOURCES, the program would called the same make-map configuration parameters and FCFs by default to reduce the raw data, and the same resultant scan maps were output.

The $\scriptsize\text{PICARD}$ recipe MOSAIC\_JCMT\_IMAGES was used to mosaic all calibrated maps pixel-by-pixel weighted by inverse-variance \citep{simpson2019}. The default WCSMOSAIC approach was chosen for this purpose. 
Next, we applied a script to remove any pixels beyond 4$\sigma$ from the median of a box region centred on each pixel. This step is necessary to prevent bad pixels from forming bright spots after matched-filtering and making them difficult to distinguish from the detected source.
Each pixel value of the mosaicked map was sampled from multiple bolometers, the root-mean-square (rms) of the bolometer values contributing to each pixel 
was the noise \citep{thomas_scuba-2}. It was contributed by the instrument and atmosphere, but which we would later call as local instrument noise \citep{simpson2019}.

The commonly used matched-filtering technique was employed to optimize source identification and characterization, and to suppress any residual large-scale noise that might not have been entirely rejected in the preceding filtering step \citep{thomas_scuba-2}. For this purpose, we requested the recipe SCUBA2\_MATCHED\_FILTER.
The matched-filtering procedure involved two steps: first, the mosaic was convolved with a wide Gaussian function with a full width at half maximum (FWHM) of 30$\arcsec$ to obtain a smooth version of the mosaicked map, which was then subtracted from the original map to remove any residual large-scale structure. Next, the image was convolved with the point spread function (PSF). For a more detailed visualization of this process, refer to \citet[][fig. 1]{chen2013a}.
Finally, after applying the matched-filtering, we obtained the SSA22 field 850 $\upmu$m map.
The flux calibration check would be presented in Section~\ref{sec:source_catalogue}.

\begin{figure}
	\includegraphics[width=\columnwidth]{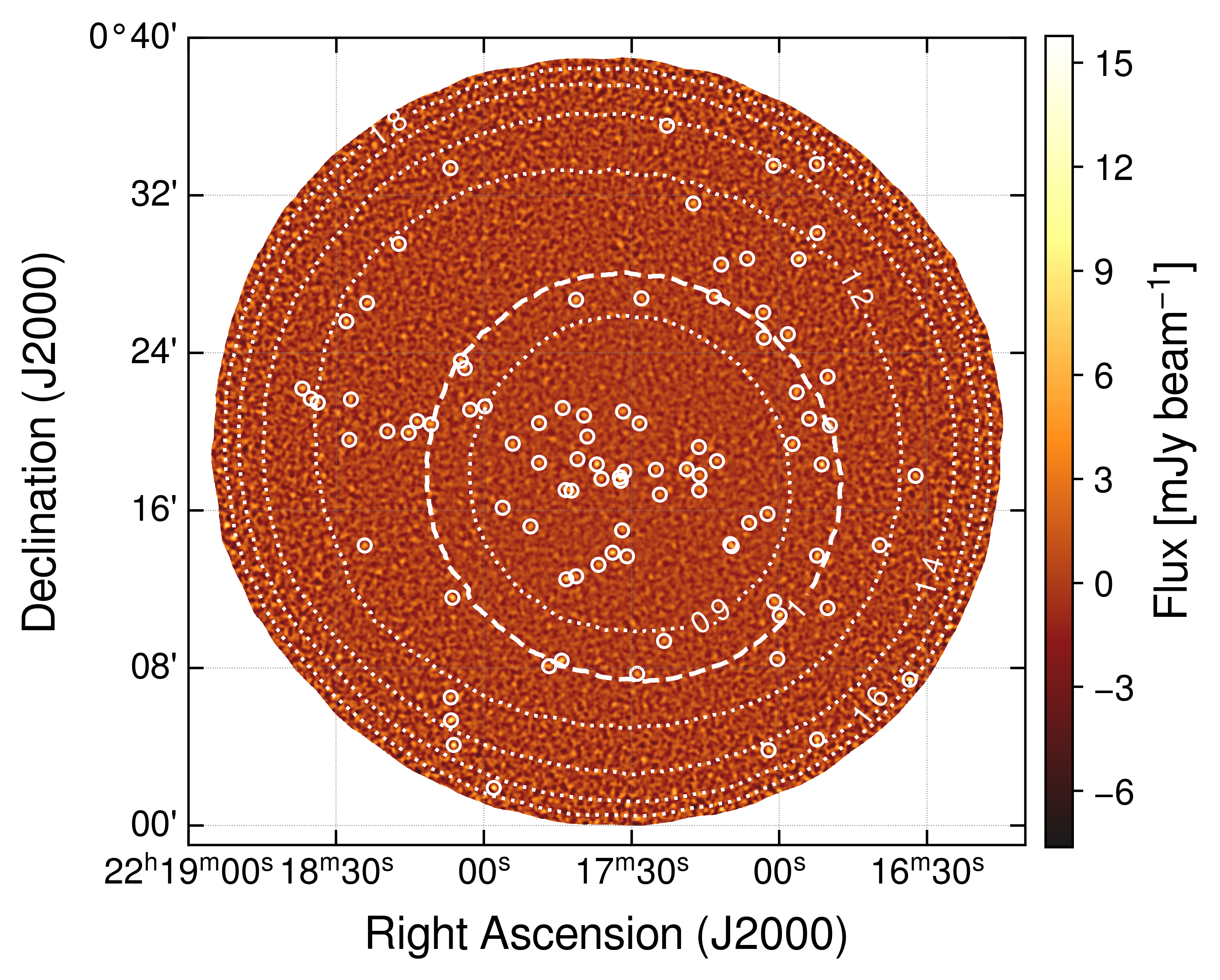}
    \caption{ 
    The presented flux density map covers an area of approximately 0.34 deg$^{2}$ with $\sigma_\text{rms}$ $\leqslant$ 2 mJy. The contours illustrate the variation in noise levels at 0.9, 1.0, 1.2, 1.4, 1.6, and 1.8 mJy, smoothly increasing from the centre outwards. The thick white dashed line delineates the region with $\sigma_\text{rms}$ $\leqslant$ 1 mJy, referred to as the `deep region' in subsequent chapters. The white circles represent the 92 sources detected above 5$\sigma$, with a circle radius equal to the FWHM of the 850 $\upmu$m empirical PSF, which measures 14$\arcsec$.}
    \label{fig:map_5sigma}
\end{figure}

\begin{figure}
	\includegraphics[width=\columnwidth]{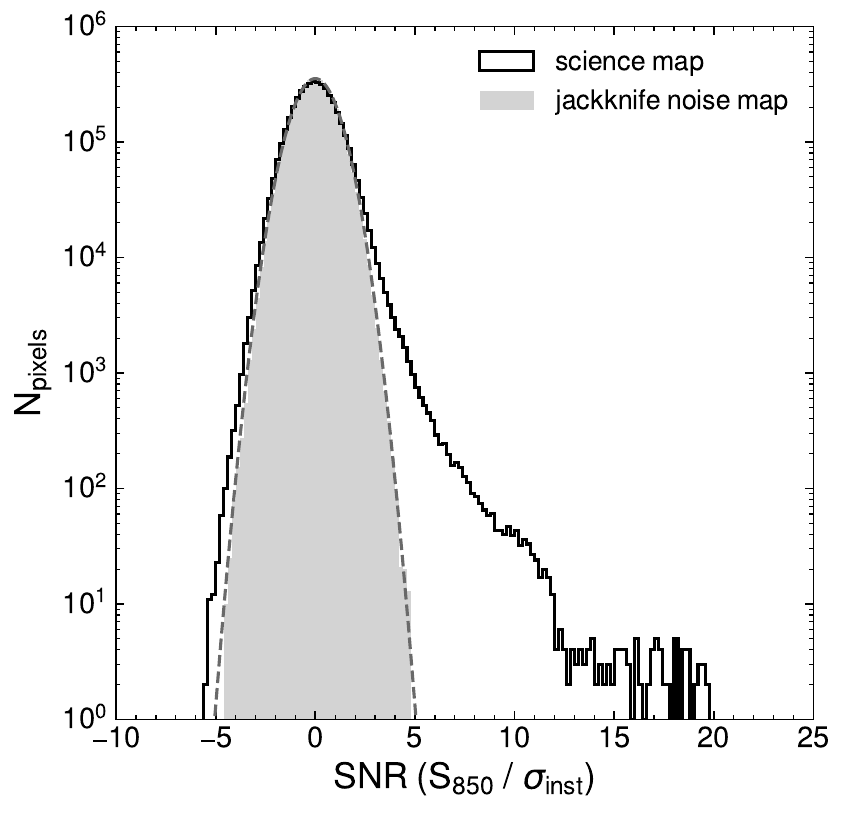}
    \caption{ 
    The pixel distribution of the signal-to-noise ratio map of the SSA22 field (black solid line). The light grey filled area corresponds to the pixel distribution of the jackknife S/N map, while the grey dashed line represents the Gaussian distribution. 
    Their good alignment indicates that the random instrument noise in the 850 $\upmu$m map follows a Gaussian distribution.
    On the other hand, the pixel distribution of the science map extends beyond the Gaussian noise on both the left and right sides. The right side reflects the submillimetre signal, while the left side results from the matched-filter process.}
    \label{fig:snr_histogram_figure}
\end{figure}

\section{Coverage map and source extraction}
\label{sec:map_extraction}

\subsection{Coverage map}
We define the effective area as the total area used for source detection and subsequent analysis, where the local instrument noise (i.e. rms) is less than 2 mJy beam$^{-1}$, corresponding to approximately 0.34 deg$^{2}$. 
The flux map of our 850 $\upmu$m mosaic is shown in Fig.~\ref{fig:map_5sigma}, the white circles represent significant sources with fluxes higher than 5 times rms, which are later used to construct the empirical point spread function. 
The white dashed lines in the figure depict the distribution of local instrument noise, due to the observation mode, the exposure time gradually decreases and the instrument noise gradually increases outward.
Within the deepest coverage of the map, approximately 5 arcmin in radius, the rms level reaches a uniform value of $\sim$ 0.79 mJy beam$^{-1}$.

In Fig.~\ref{fig:snr_histogram_figure}, we present the pixel value histogram of the signal-to-noise ratio (S/N) map, which exhibits excess in both positive and negative directions relative to Gaussian noise (indicated by the grey dashed line or light grey shadow). 
The long tail of the S/N distribution, extending almost to 20, is shaped by the emission of astrophysical objects. The negative excess is contributed by the negative rings (see Section~\ref{sec:model_psf}) around the source peaks introduced by the matched-filter in the final data reduction step.

\subsection{Source extraction}

\subsubsection{Modelling PSF}
\label{sec:model_psf}
The SCUBA-2 instrumental response to 850 $\upmu$m light from distant galaxies can be expressed by a two-component Gaussian function, with an expected effective FWHM of $\sim$ 12$\farcs$6 \citep{mairs2021}. However, during the matched-filtering reduction, the shape of the instrumental PSF undergoes changes. As a result, it becomes necessary to reconstruct the PSF to accurately extract the filtered submillimetre sources.

\begin{figure}
	\includegraphics[width=\columnwidth]{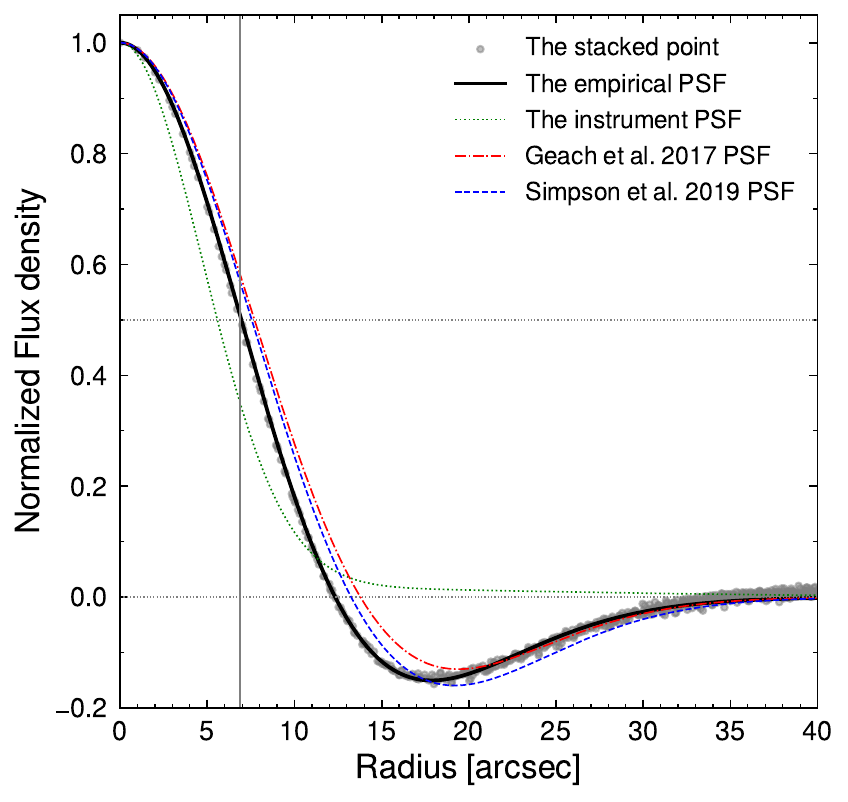}
    \caption{ 
    The profile of the reconstructed PSF model for the 850 $\upmu$m mosaic map. The grey dots represent the normalized median profile of all sources with significance greater than 5$\sigma$. The black solid line depicts the best fitting curve of the empirical PSF, with the thick grey vertical line flagging the half of FWHM of $\sim$ 6$\farcs$95. 
    The green dotted curve represents the instrumental PSF profile from \citet{mairs2021}. The circular depression around the summit of the source arises from the matched-filter, which has been updated \citep{mairs2021}, resulting in a slimmer empirical PSF compared to previous literature.}
    \label{fig:empirical_PSF_figure}
\end{figure}

To reconstruct the PSF, we first identified all sources with significance greater than 5$\sigma$ over the effective area and extracted the pixel values within a radius of 40$\arcsec$ of each source. Subsequently, we stacked the pixel values of every source and obtained the median point spread function. The reshaped profile of the 850 $\upmu$m sources, with a FWHM of approximately 14$\arcsec$, is depicted in Fig.~\ref{fig:empirical_PSF_figure}. 
This profile exhibits a circular depression resulting from the application of the matched-filter to the map and instrumental PSF. Notably, the width of the empirical PSF is slightly narrower than previously reported due to updates in the matched-filter \citep{mairs2021}.

The result of the reconstruction can be effectively described by a combination of two Gaussian functions, as presented in previous literature \citep{geach2017, simpson2019}:
\begin{equation}
    P(\theta)= A_1 \times \exp( \frac{-\theta^2}{2 \times \sigma_1^2}) 
    - A_2 \times \exp( \frac{-\theta^2}{2 \times \sigma_2^2})
    \label{eq:emprical_PSF}
\end{equation}
with the best fitting values are $A_1$ = 2.07, $A_2$ = 1.07, $\sigma_1$ = 7$\farcs$71 and $\sigma_2$ = 11$\farcs$08.

\begin{figure*}
	\includegraphics[width=2\columnwidth]{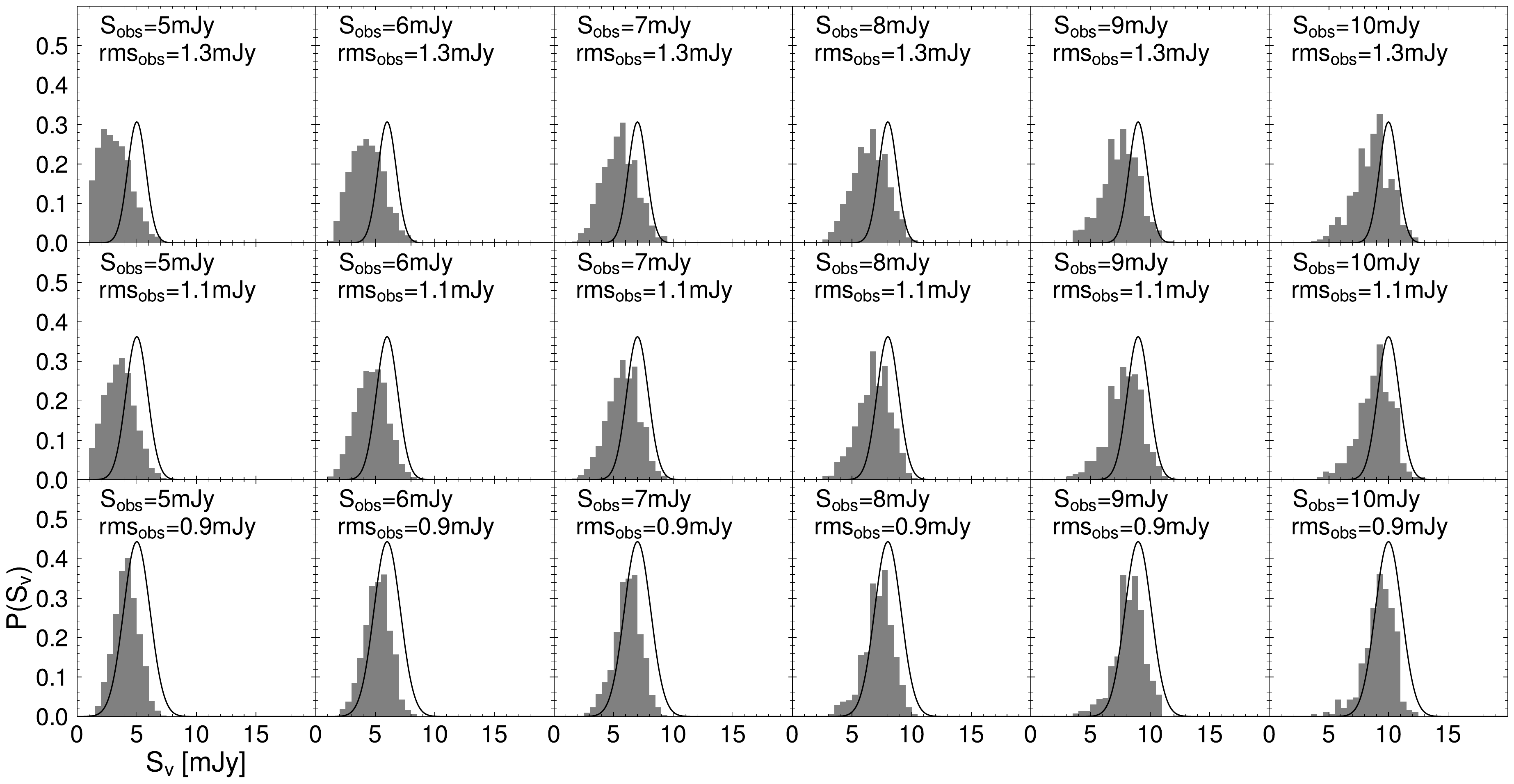}
    \caption{ 
    An example of the injected (intrinsic) flux density probability distribution p($S_\text{true}$) (grey shaded bars) corresponding to the two-dimensional bin $(S_\text{obs},\,\sigma)$, obtained through the empirical recovery method from 5000 simulations. The black solid line represents the observed flux distribution, assuming a Gaussian distribution centred on the observed flux, with the standard deviation being the local instrument noise. The observed flux increases from left to right, and the rms decreases from top to bottom, with their values shown in the upper left corner of each stamp. As the observed flux increases and the local instrument noise decreases, the effect of flux boosting becomes less significant.}
    \label{fig:observed_flux_distribution_figure}
\end{figure*}

\subsubsection{Source extraction}
\label{sec:source_extraction}
Dusty submillimetre galaxies in the early universe typically occupy compact regions \citep[with extents of a few kiloparsecs;][]{umehata2017}. Due to their small angular sizes compared to the SCUBA-2 average beam size \citep[approximately 130 kpc at the typical redshift of $\sim$ 2-3 of SMGs;][]{hayward2013a,lovell2021}, these distant dust emission populations appear as unresolved point sources in SCUBA-2 survey maps \citep{casey2013,shim2020,zhang2022}.
In addition, the data reduction pipeline uses the default units of mJy beam$^{-1}$, meaning that the pixel value of each source peak in the map represents its total measured submillimetre flux density (see Section~\ref{sec:data_reduction} for context on flux calibration).

The matched-filter fully optimizes the science map for source extraction and enhances the clarity of each source. For detection and source identification, we adopted a simple yet common `top-down' iterative algorithm, following the approach used in \citet{wang2004,wang2017}. 
Firstly, we located the most prominent peak in the signal-to-noise ratio map and subtracted 5 per cent \citep[referred to as `CLEAN gain';][]{wang2017} of the peak flux scaled PSF from a circular region with a radius of 40$\arcsec$, centred on this peak in the flux density map. 
We then recorded the noise level and the subtracted flux density, along with the peak pixel position, which served as the location of the detected source. 
This process was repeated, identifying the most significant peak, subtracting part of the peak flux, and recording the source information until the signal-to-noise ratio peak reached a predetermined detection threshold, set to 3$\sigma$ to ensure the adequate detection of potential sources.

When identifying a peak within a radius of 7$\arcsec$ (approximately half FWHM of the PSF) of a previously detected source, we considered them as a single source and continued subtracting at the original extracted location. Once the iterative procedure was completed, we combined each piece of cleaned flux density with the remaining flux below the threshold at the identified location to obtain the total flux density for each source.
The final catalogue consisted of 390 sources selected from the primary catalogue, with a cut limit of 3.5$\sigma$. While the choice of 3.5$\sigma$ was somewhat arbitrary, it was commonly used in the literature, and the total false sources accounted for 17 per cent (see Fig.~\ref{fig:FDR_figure}).

In the literature, the commonly adopted proportion for peak subtraction was 100 per cent. However, this method sometimes produced fake sources \citep{zhang2022}, which was also observed in our tests. The `CLEAN' method, involving a gradual deduction of peak flux, effectively separated somewhat blended sources. Comparing to the catalogue from \citet{geach2017}, this subtraction approach allowed us to distinguish two pairs of close submillimetre sources (with separations less than the FWHM). Besides, our method identified four additional pairs of close sources in the 3.5$\sigma$ cut-off catalogue, which could be valuable for follow-up observations and the search for counterparts.

\section{THE MONTE CARLO SIMULATION}
\label{sec:simulation}

Because of the random measurement errors of the instrument during the reception of submillimetre photons, a certain luminosity of light can be `scattered' to higher or lower values, resulting in a Gaussian or Poisson distribution. 
Moreover, within the detection capability of telescope, there are always more faint sources than bright ones. This situation leads to a higher number of events of dimmer sources being erroneously assigned higher flux densities during observations than the opposite case, presenting a persistent risk of flux overestimation.

The phenomenon of flux overestimation, also known as `flux boosting', is influenced by both the Eddington and Malmquist biases discussed above. Also, source blending of multiplicity can contribute to the boosting effect. 
On the other hand, some of the faintest sources may be undetected due to being obscured by random noise, while some noise peaks may be misidentified for the sources because of random fluctuates of noise.
Moreover, the relatively large FWHM of the SCUBA-2 beam, coupled with the pointing accuracy of JCMT of around 1-2 arcsec, introduces a considerable level of uncertainty in the positional accuracy of detected sources.

Given the challenges in accurately recognizing and extracting sources, conducting Monte Carlo simulations becomes essential to obtain a robust catalogue and reliable number counts, and facilitate follow-up observations of individual sources.

\begin{figure*}
	\includegraphics[width=2\columnwidth]{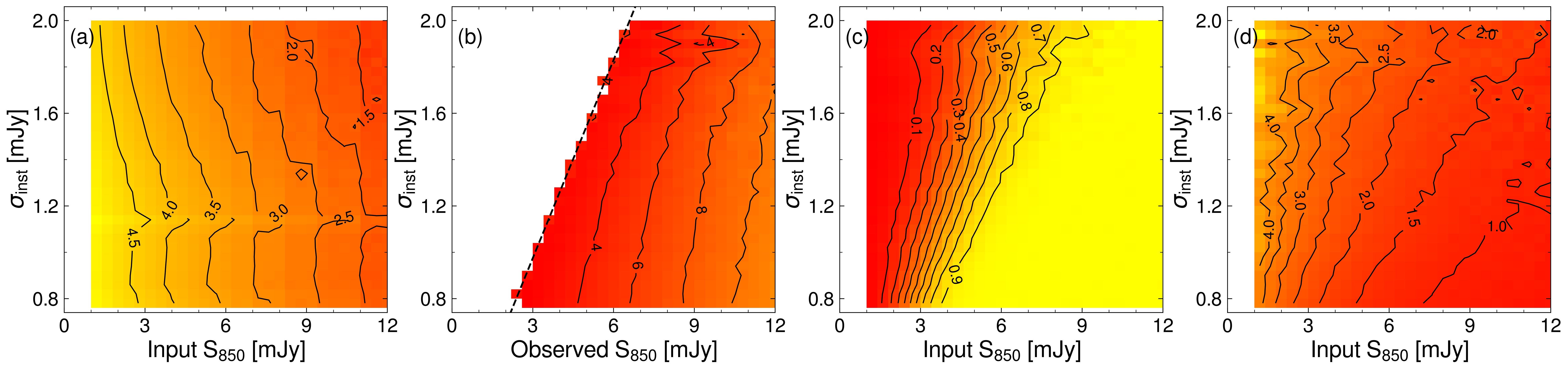}  
    \caption{
    The graphical statistical outcomes from 5000 Monte Carlo simulations.
    Panel (a) illustrates the number distribution of injected sources, written in log$_{10}(N)$. This distribution provides insights into the mapping sensitivity in relation to the area covered. The presence of horizontal spurs with rms values of approximately 0.8 and 1.15 mJy suggests slightly larger covered areas.
    Panel (b) presents the input flux density distribution as a function of the output flux density and noise, which is used for deboosting each source with a significance of $\geqslant$ 3.5$\sigma$ in the science map.
    Panel (c) represents the completeness of the source extraction, which is related to the input flux density and the instrument noise. It is calculated by dividing the observed number of artificial sources by the total number of sources injected into the jackknife map.
    Panel (d) displays the positional error between the extracted position and the actual location of the 850 $\upmu$m source, and is a function of the input flux density and the local noise.}
    \label{fig:heatmap_row}
\end{figure*}

\subsection{Simulation Method}
\label{sec:simulation_method}

First, we needed a jackknife map. We randomly divided the total of 134 calibrated scan maps into two groups. Each group consisted of half of the new data and half of the archived data, ensuring a balanced distribution of exposure time between the two subsets. Subsequently, we synthesised two half-mosaics separately and performed a subtraction between them based on the world coordinate system. The pixel values in the resulting subtracted map were then scaled using the formula $\sqrt{t_1 \times t_2}/(t_1+t_2)$, where $t_1$ and $t_2$ represented the exposure times of the corresponding pixels in the two half-mosaics \citep{chen2013a}. After applying additional steps to remove bad pixels and use matched-filter, the resulting map was designated as the jackknife map, also known as the true noise map \citep{chen2013a, wang2017}.
Since the astrophysical signals from the two half-mosaics were eliminated during the subtraction process, the flux layer of the jackknife map contained only Gaussian instrument noise (as indicated by the grey-filled portion in Fig.~\ref{fig:snr_histogram_figure}). Thus, this map provided a sophisticated estimation of the true random noise in the science flux density map. The jackknife rms map showed a high level of consistency with the science rms map \citep{wang2017}.

In the second step, we conducted a Monte Carlo simulation by injecting millions of artificial sources into our jackknife true noise map. As the fundamental model, we adopted the differential counts from \citet{geach2017}, represented by a Schechter function:
\begin{equation}
    \frac{\text{d}N}{\text{d}S} = 
    \left( \frac{N_0}{S_0} \right) 
    \left( \frac{S}  {S_0} \right)^{-\gamma} 
    \exp \left( -\frac{S}{S_0} \right).
    \label{eq:Schether_form}
\end{equation}
Considering that the observed flux densities of the brightest and faintest sources detected in the science 850 $\upmu$m map were approximately 3 and 15 mJy beam$^{-1}$, respectively, we set the flux density of the injected artificial sources to range from 1 to 20 mJy. \citet{hayward2013b} indicated that this faintest flux setting below the detection limits of almost single-dish surveys. We divided this injected flux range into 20 bins in logarithmic space and calculated the theoretical number of sources in each flux density bin according to the counts model. For simulated flux densities greater than $\sim$ 13 mJy, the small map area resulted in theoretical source numbers less than one. 
To ensure sufficient sources in the bright flux density bin and obtain accurate statistics, we applied Poisson randomisation to determine the actual number of simulated sources in each bin. The flux density of each artificial source was then randomly and uniformly sampled from the corresponding flux bin.

For injecting the artificial sources into the jackknife map, we utilized the PSF outline of the filtered sources obtained in Section~\ref{sec:model_psf}. Each source was placed randomly at any pixel in the map, regardless of any clustering effects \citep{simpson2019}, and anywhere within the pixel. After injection, we performed the source extraction program described in Section~\ref{sec:source_extraction} on the simulation map, repeating the injection and extraction procedure 5000 times for a robust statistical analysis.

Finally, we matched the injection and extraction source catalogues. In each simulation, if a source with a significance of $\geqslant$ 3.5$\sigma$ from the extracted catalogue was found within a radius of 0.75 times the FWHM (i.e. 0.75 $\times$ 14 arcsec $\sim$ 11 arcsec) of an injected source, we considered the injected source as `recovered'. In cases where an extraction matched multiple injections, we retained the brightest component for the match. 
Once an injected source was successfully matched, it was excluded from further matching attempts. Sources from the injection catalogue that remained unmatched were considered to be submerged in noise and not extracted, while unmatched sources from the extraction catalogue were regarded as false sources resulting from Gaussian noise.

\begin{figure*}
	\includegraphics[width=2\columnwidth]{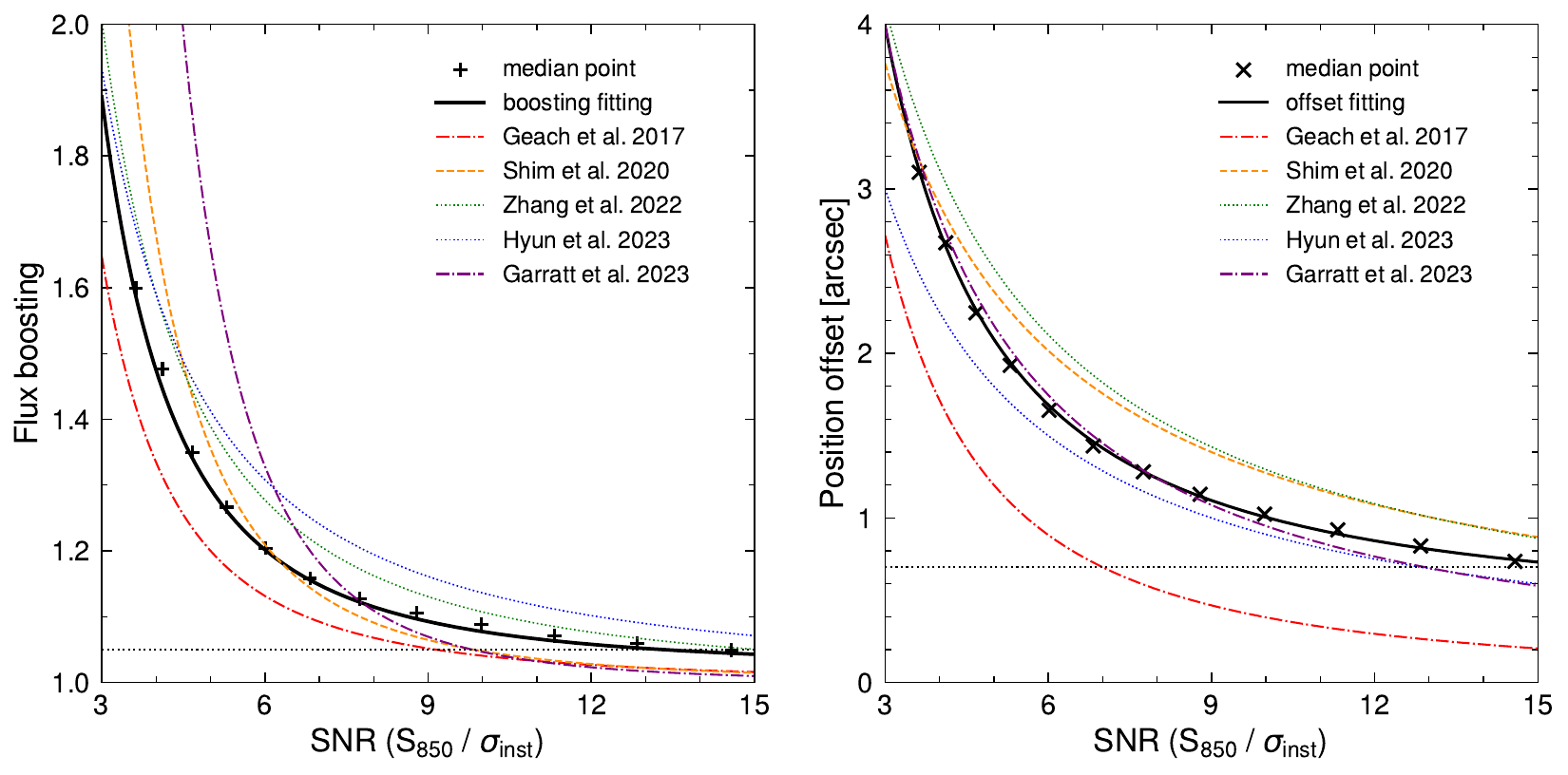}
    \caption{ 
    (Left image) The median flux boosting is shown as a power law function of measured S/N, representing the ratio between recovered and intrinsic flux density of the same source. 
    The `+' markers symbolise the median boosting result from the simulation, and the black solid line is the best-fitting curve. The horizontal black dotted line marks the value of 1.05. The boosting curve shows a significant drop within the range of S/N < 6, followed by a gentle decline to about 1.04. Overall, the boosting effect in our map falls in between other studies.
    (Right image) The positional uncertainty is defined as the distance between the extracted position and the input position of the same source. The `x' markers represent the median offset result, and the best-fitting of positional error is depicted in a power-law form (black solid curve). The horizontal black dotted line indicates the value of 0.7. The positional error decreases steadily from 3.3 to 0.73 arcsec with the increase of observed S/N.}
    \label{fig:boosting_offset_powerlaw_figure}
\end{figure*}

\subsection{Flux boosting}
\label{sec:flux_boosting}
To estimate the flux boosting, we compared the extracted and injected flux densities of each matched source, which depended on both the observed flux density and the local instrument noise. As sources with the same injected flux were affected by different local noise perturbations, their observed fluxes could vary, in turn, an observed flux has a corresponding intrinsic flux density probability distribution, denoted as p($S_\text{true}$).

Following an empirical method \citep{geach2017}, we obtained p($S_\text{true}$) for each two-dimensional bin $(S_\text{obs},\,\sigma)$ by calculating the histogram of flux density of injected sources falling into that bin. This allowed us to estimate the intrinsic (deboosted) flux density of the detected sources in the real map. The median value of the injected flux distribution was used as the deboosted flux density of the source, and the standard deviation was considered as the uncertainty in the deboosting process \citep{shim2020, zhang2022}. 
An instance of p($S_\text{true}$) is shown in Fig.~\ref{fig:observed_flux_distribution_figure}, where the centre of the injected flux density distribution gets closer to the observed flux as the observed flux increases, indicating a decreasing flux boosting effect with increasing observed flux density. Similarly, as the local noise decreases, the flux boosting becomes weaker.

Once the matching procedure of the 5000 simulated input and output catalogues was completed, we aggregated all two-dimensional bins and the relevant median results of the injected flux density distribution to create a two-dimensional image (Fig.~\ref{fig:heatmap_row}, panel (b)), which was used for deboosting the sources extracted from the SSA22 850 $\upmu$m map. 
Meanwhile, we summarized the statistical results regarding the injected source number, completeness, and positional uncertainty using a flux interval of 0.4 mJy and an rms interval of 0.04 mJy.

We used the meshed values of the panel as a look-up table to estimate the correction for each source, based on their observed flux density and rms. By locating the corresponding $(S_\text{obs}, \sigma)$ bin for observed flux density and instrument noise of a source, we derived the deboosted flux and its uncertainty, which, combined with the instrument noise and confusion noise, provided the total uncertainty of the source in the real catalogue \citep{shim2020}.

The average flux boosting effect could be described by a simple power-law function of the measured signal-to-noise ratios of sources \citep{geach2017,simpson2019,shim2020}:
\begin{equation}
    \mathcal{B} = 1.02 + 0.15 \times  ( \frac{\text{SNR}}{6.53} ) ^{-2.26}.
    \label{eq:flux_boosting}
\end{equation}
In Fig.~\ref{fig:boosting_offset_powerlaw_figure}, the average flux boosting reaches an amplification factor of 1.63 for 3.5$\sigma$ sources, and the boosting curve decreases to about 1.04 at a S/N level of 15, with the average boosting factor for all sources being about 1.4. It is essential to note that this power-law curve is an approximation, and the flux boosting of each source depends on its observed flux density and local instrument noise, rather than being directly derived from the above equation~(\ref{eq:flux_boosting}).

We conducted a simple comparison of flux boosting with results from literature. In the region where the S/N is significant, our flux boosting is roughly consistent with \citet{zhang2022}, slightly higher than \citet{geach2017}, \citet{shim2022}, and \citet{garratt2023}, and slightly lower than \citet{hyun2023}. We attribute these differences primarily to variations in the uniformity and coverage areas of surveys. Additionally, the presence of denser regions, where faint objects below the detection threshold crowd into the instrument beam, can lead to a higher boosting curve. 
The higher boosting values in the low S/N section reported by \citet{shim2022} and \citet{garratt2023} can likely be attributed to the higher noise levels of their maps. Differences in simulation performance may also contribute to the distinct variations in the boosting results \citep{scott2008, scott2010, wang2017, zavala2017}.

\begin{figure}
	\includegraphics[width=\columnwidth]{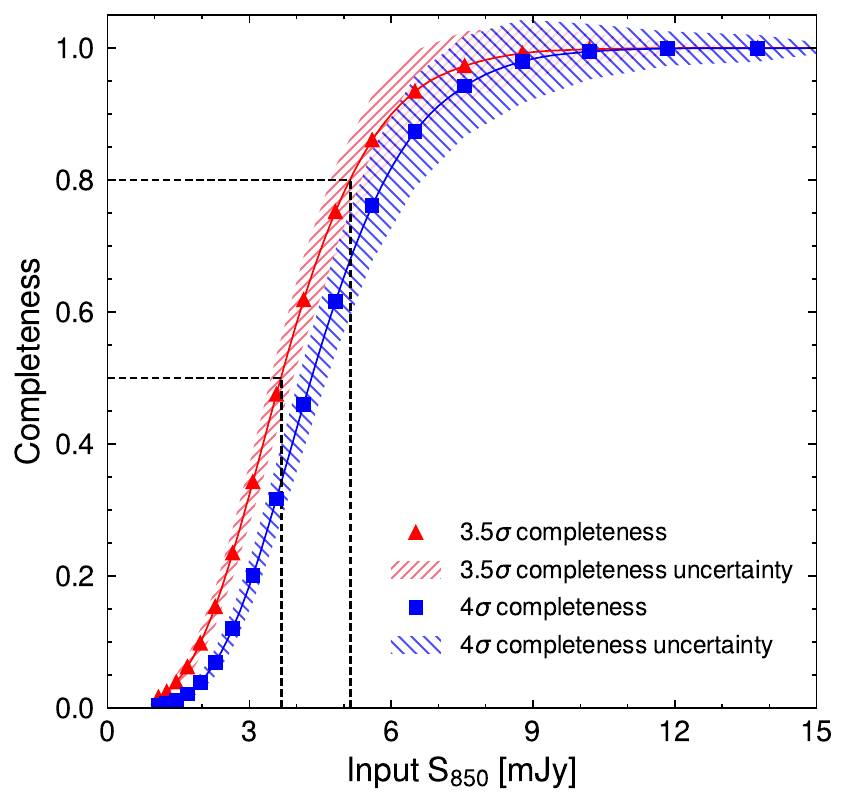}
    \caption{ 
    The average completeness as a function of input flux density, is obtained through comparing the recovered sources number to the number of injected artificial sources. 
    The triangle markers symbolize the average completeness at 3.5$\sigma$, the corresponding values of 50 and 80 per cent is 3.7 and 5.1 mJy, respectively. 
    In this work, we only analyse the sources of 3.5$\sigma$ and above, the completeness of 4$\sigma$ displayed in graph is for comparison and shows the change in completeness under different measurement significance.
    }
    \label{fig:average_completeness_figure}
\end{figure}

\subsection{Completeness}
Completeness is defined as the ratio of the number of recovered sources to the number of injected sources in the input flux density and instrument noise bin.
To estimate the completeness correction for each source in the science map and the randomly sampled flux density of sources in Section~\ref{sec:number_counts}, we used the two-dimensional meshed image (Fig.~\ref{fig:heatmap_row}, image (c)).
From the figure, it is evident that the efficiency of extracting simulated sources rapidly decreases when the input flux density is lower (< 8 mJy) due to the increase in rms, indicating that noise significantly influences the detection rate of sources in the low flux region.

The 50 per cent completeness limit corresponds to 2.6 mJy in the deepest region of the map and 6.2 mJy in the shallowest region. Similarly, the 80 per cent completeness limit corresponds to 3.4 and 8 mJy, respectively. It is essential to note that these corresponding flux values are the deboosted (intrinsic) flux densities of the sources. The average completeness shown in Fig.~\ref{fig:average_completeness_figure} has a simple relationship with the input flux density, achieving 50 per cent and 80 per cent completeness when the input flux density is 3.7 and 5.1 mJy, respectively.

\subsection{Positional uncertainty}
We also determined the positional uncertainties of the detected sources using a similar method. The positional uncertainty of a source is correlated with its injected flux density and local noise. 
The median positional errors are shown in graph (d) of Fig.~\ref{fig:heatmap_row} and can be described by a mathematical form similar to boosting:
\begin{equation}
    \delta\theta = 0\farcs40 + 1\farcs83\times(\frac{\text{SNR}}{4.73})^{-1.48}.
    \label{eq:position_offset}
\end{equation}
The positional offset gently decreases from 3.3 to 0.73 arcsec within the observed S/N range of 3.5 to 15, the position uncertainty for all sources is about 2.5 arcsec on average.
The positional accuracy is usually influenced by the mapping depth, uniformity, and the survey coverage area. 
When using the positions of the SMGs detected in the 850 $\upmu$m map for follow-up observations in other wavebands, caution is advised due to both the telescope pointing accuracy and the position uncertainty of the source extraction, which can reach about a few arcsec.

\begin{figure}
	\includegraphics[width=\columnwidth]{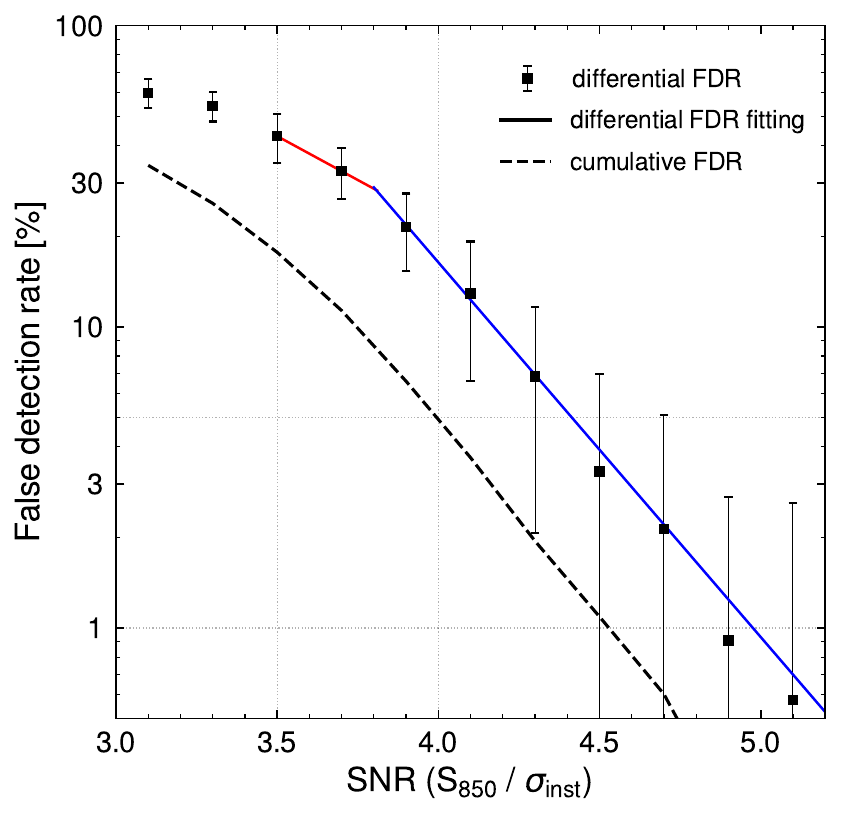}
    \caption{ 
    The false detection rate is defined as the ratio of the number of `sources' detected in the random noise of the jackknife map to the number of detected sources on the science map. 
    The red and blue solid lines represent the best-fitting lines for the differential FDR, corresponding to two linear functions in logarithmic space. 
    At the detection significance level of 3.5$\sigma$, the differential and cumulative FDRs are 43 and 17 per cent, respectively. Notably, the differential FDR sharply decreases to zero as the source detection significance reaches 5.2$\sigma$.}
    \label{fig:FDR_figure}
\end{figure}

\subsection{False detection rate}
As mentioned earlier, fluctuations of Gaussian noise could occasionally be misidentified as sources. To measure the false detection rate (FDR), we generated 40 jackknife true noise maps, and compared the number of fake sources detected in these maps to the number of extracted sources in the real map within each S/N bin.

The signal-to-noise ratio pixel histogram (Fig.~\ref{fig:snr_histogram_figure}) illustrates that noise still contributes significantly at 3.5$\sigma$, accounting for up to 43 per cent of the detected sources (see Fig.~\ref{fig:FDR_figure}). At 4.5$\sigma$, the differential false detection rate is about 4 per cent, and when the source significance reaches 5$\sigma$, the FDR drops to approximately 1 per cent. Therefore, caution is warranted when considering sources with low detection significance.

Despite the presence of a large number of false sources at low detection significance, we chose an extraction threshold of 3.5$\sigma$ to achieve a more comprehensive source catalogue, as most of the sources were genuine. The cumulative false detection rate for sources with S/N > 3.5$\sigma$ was 17 per cent, meaning that out of all 390 sources, about 66 sources were false. Imitating the approach of \citet{geach2017}, we fitted the differential FDR with a piecewise linear function in logarithmic space:
\begin{equation}
    \text{1og}_{\text{10}}(\mathcal{F}) = \begin{cases}
        1.670 - 0.582 \times \text{SNR}, & \text{SNR}\leqslant 3.8 \\
        4.190  -1.244 \times \text{SNR}, & \text{SNR} > 3.8
        \end{cases},
    \label{eq:FDR}
\end{equation}
then we applied the fitting function to determine the differential FDR correction for each observed source.

An alternative approach to evaluate the FDR involved directly dividing the number of unmatched sources in the output catalogues of simulation by the number of sources extracted in the science 850 $\upmu$m map within each S/N bin \citep{wang2017}. 
The FDR results obtained from both methods were highly consistent since they were based on the same underlying principle: the unmatched sources primarily originated from Gaussian noise fluctuations in the jackknife map.

\begin{table*}
	\caption{The 850 $\upmu$m catalogue contains 390 submillimetre sources greater than 3.5$\sigma$. Among these, 92 detections achieve a significance of 5$\sigma$. Below are the top 10 most remarkable sources from the catalogue, sorted by significance.
	The column header S$_\text{850}^\text{obs}$ $\pm$ $\sigma_\text{inst}$ is the measured flux density and the local instrument noise. 
	The column S$_\text{850}^\text{deb}$ $\pm$ $\sigma_\text{tot}$ represents the deboosted flux density and the total uncertainty, which is consisted of the deboosting uncertainty, instrument noise and confusion noise. 
	The columns $\mathcal{C}$, log$_{10}\mathcal{F}$, $\delta\theta$ give the source completeness, differential FDR in logarithmic form and position offset, respectively. The full catalogue is available in the online journal.}
	\begin{threeparttable}
	\begin{tabular}{cccc r@{ $\pm$ }l c r@{ $\pm$ }l ccc}

	\hline
	SSA22 source name & SSA22 ID & R.A.(J2000) & Dec.(J2000) & \multicolumn{2}{c}{S$_\text{850}^\text{obs}$ $\pm$ $\sigma_\text{inst}$} & S/N & \multicolumn{2}{c}{S$_\text{850}^\text{deb}$ $\pm$ $\sigma_\text{tot}$} & $\mathcal{C}$ & log$_{10}\mathcal{F}$ & $\delta\theta$ \\
	
	& & (hms) & (dms) & \multicolumn{2}{c}{(mJy)} & & \multicolumn{2}{c}{(mJy)} & & & (arcsec) \\
	
	\hline
	SSA22\_J221732+001740  &  SSA22.0000  &  22:17:32.37  &  +00:17:40.70  &  15.7 &  0.8  &   19.8  &  15.0 &  2.2  &  1.00  &  -20.41  &  0.64 \\
	SSA22\_J221718+001804  &  SSA22.0001  &  22:17:18.77  &  +00:18:04.70  &  9.6  &  0.8  &   12.0  &  9.2  &  1.5  &  1.00  &  -10.71  &  0.79 \\
	SSA22\_J221737+001819  &  SSA22.0002  &  22:17:37.03  &  +00:18:19.70  &  9.3  &  0.8  &   11.9  &  8.7  &  1.4  &  0.99  &  -10.66  &  0.85 \\
	SSA22\_J221659+001039  &  SSA22.0003  &  22:16:59.90  &  +00:10:39.70  &  11.6 &  1.0  &   11.9  &  10.8 &  1.8  &  1.00  &  -10.59  &  0.82 \\
	SSA22\_J221733+001350  &  SSA22.0004  &  22:17:33.83  &  +00:13:50.70  &  8.9  &  0.8  &   11.2  &  8.2  &  1.5  &  1.00  &  -9.77   &  0.93 \\
	SSA22\_J221742+001659  &  SSA22.0005  &  22:17:42.17  &  +00:16:59.70  &  9.6  &  0.8  &   11.2  &  9.2  &  1.5  &  1.00  &  -9.72   &  0.79 \\
	SSA22\_J221743+001229  &  SSA22.0006  &  22:17:43.23  &  +00:12:29.70  &  9.0  &  0.8  &   10.7  &  8.7  &  1.5  &  1.00  &  -9.18   &  0.90 \\
	SSA22\_J221651+001819  &  SSA22.0007  &  22:16:51.37  &  +00:18:19.70  &  10.2 &  1.0  &   10.5  &  9.2  &  1.6  &  1.00  &  -8.89   &  0.97 \\
	SSA22\_J221656+002844  &  SSA22.0008  &  22:16:56.03  &  +00:28:44.70  &  11.7 &  1.2  &   10.1  &  10.6 &  2.0  &  1.00  &  -8.33   &  0.98 \\
	SSA22\_J221728+002024  &  SSA22.0009  &  22:17:28.37  &  +00:20:24.70  &  7.3  &  0.8  &    9.2  &  6.7  &  1.4  &  0.99  &  -7.22   &  1.04 \\
	$\cdots$ & $\cdots$ & $\cdots$ & $\cdots$ & \multicolumn{2}{c}{$\cdots$} & $\cdots$ & \multicolumn{2}{c}{$\cdots$} & $\cdots$ & $\cdots$ & $\cdots$ \\
	\hline
	
	\end{tabular}
	\label{tab:source_cata}
	
	\end{threeparttable}

\end{table*}

\subsection{Source catalogue} 
\subsubsection{Confusion noise} 
Confusion noise is a persistent challenge in astronomical measurements \citep{scheuer1957, condon1974, takeuchi2004}, arising from the combined effects of thermal emission from Galactic dust and crowding by unresolved faint extragalactic sources \citep{helou_beichman1990}. In the context of this paper, cirrus confusion is negligible, and thus, it will not be considered further. 
Nevertheless, confusion noise remains an unmitigated factor, impervious to longer exposure times or increased detector sensitivity \citep{helou_beichman1990}, affecting both position and photometry measurements \citep{hogg2001}. While not always rigorously accounted for in the SCUBA-2 signal map, we provide a simple estimation of its impact.

A commonly used rule of thumb for estimating confusion noise is that the confusion limit is reached when the source density reaches approximately one source per 20--30 resolution elements \citep{hogg2001, dole2003}. Here, a resolution element is the beam area defined as $\Omega_\text{beam} = \pi \sigma^2$, with $\sigma$ $\approx$ FWHM/2.35 for the Gaussian assumption \citep{hogg2001}. 
Following \citet{hogg2001} and utilizing the final differential counts (Section~\ref{sec:number_counts}), the calculated confusion limit is approximately 1.49 mJy, indicating that the source density reaches one per 30 beams when the flux is below 1.49 mJy. Consequently, at our 3.5$\sigma$ threshold, the estimated confusion noise is $\sim$ 0.43 mJy beam$^{-1}$ \citep{dole2003, simpson2019}.

We followed \citet{helou_beichman1990} and utilized the equation:
\begin{equation}
    \sigma^2_\text{c}(S_\text{lim}) = \Omega_\text{bm} \int^{S_\text{lim}}_{0} S^2 \frac{\text{d}N(S)}{\text{d}S} \text{d}S,
    \label{eq:helou&beichman1990}
\end{equation}
where $\Omega_\text{bm}$ represented the 850 $\upmu$m beam area of SCUBA-2 with a value of 242 arcsec$^2$ \citep{geach2017}, and S$_\text{lim}$ was the confusion limit flux of 1.49 mJy given by the rule of thumb above. Based on our final number counts, the estimated confusion noise was $\sim$ 0.50 mJy beam$^{-1}$, consistent with former.
If we considered the equation~(\ref{eq:helou&beichman1990}) with S$_\text{lim}$ set to infinity, the derived value of 0.95 mJy was close to the confusion noise of 0.86 mJy proposed by \citet{geach2017}, but this value was influenced by bright sources \citep{cowie2017, geach2017}.

\citet{cowie2017} highlighted that the signal-to-noise ratio of the signal residual map exhibits excess over the Gaussian instrument noise due to confusion noise \citep[see][fig. 3(a)]{cowie2017}. 
For a test, we used the equation from \citet{cowie2017} and \citet{simpson2019} as follows:
\begin{equation}
	\sigma_\text{c}=\sqrt{\sigma_\text{total}^2-\sigma_\text{stat}^2},
    \label{eq: cowie2017}
\end{equation}
where $\sigma_\text{total}$ was the total measured noise, encompassing both confusion noise $\sigma_\text{c}$ and statistical noise $\sigma_\text{stat}$ from the signal map \citep{cowie2017}. 
Here, $\sigma_\text{total}$ and $\sigma_\text{stat}$ represented the dispersion of the signal subtracted map excluded the 3.5$\sigma$ detected sources and the jackknife true noise map, respectively \citep{simpson2019}. 
The jackknifing process removed the bright sources from the astronomical signal map, as well as the confusion noise caused by faint sources below the detection limit \citep{austermann2010}, thus $\sigma_\text{stat}$ came from Gaussian instrument noise indeed.
Due to the dominance of instrument noise throughout the SCUBA-2 850 $\upmu$m map and its non-uniform nature, increasing with radius due to the observation pattern \citep{simpson2019}, we calculated the value only in the deepest region with $\sigma_\text{rms}$ $\leqslant$ 1 mJy beam$^{-1}$. The resulting value of $\sim$ 0.28 mJy beam$^{-1}$ was in basic agreement with previous estimates, the confusion noise would be $\sim$ 0.30 mJy beam$^{-1}$ when the entire map was considered in the calculation process.

Because of the description above, we estimated the confusion noise to be 0.43 mJy beam$^{-1}$. The total uncertainty of the deboosted flux density for each source is obtained as the square root of the sum of squares of confusion noise, instrument noise, and deboosting error \citep{shim2020, hyun2023}.

\begin{figure}
	\includegraphics[width=\columnwidth]{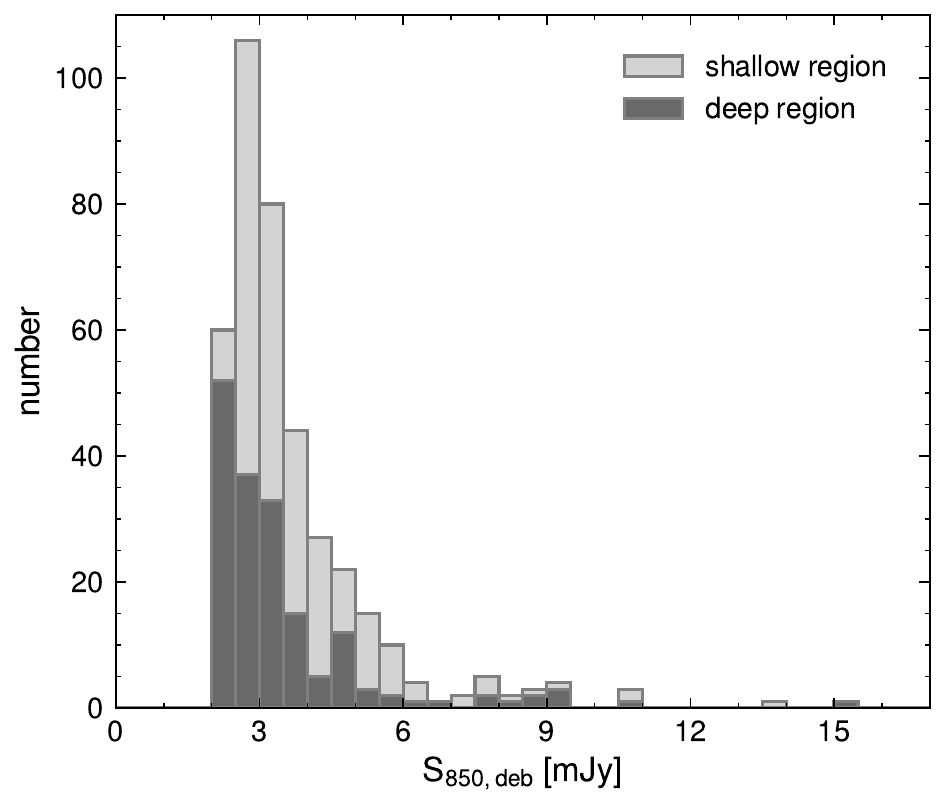}
    \caption{ 
    The number distribution of the 850 $\upmu$m deboosted flux density for the 390 detected SCUBA-2 sources great than 3.5$\sigma$. The dark grey bars represent the flux density distribution of 171 sources in the deeper region, where the rms is $\leqslant$ 1 mJy in the astronomical map. On the other hand, the light grey bars display the flux density distribution of 219 sources in the shallower region.}

    \label{fig:number_distribution}

\end{figure}

\subsubsection{Source catalogue}
\label{sec:source_catalogue}
We conducted a cross-match between the source catalogue with the S2CLS survey \citep{geach2017}, and checked both the observed and deboosted fluxes of the matched sources. The sources in our catalogue with a significance greater than 4$\sigma$ were used for matching, with a matching radius of 7", resulting in a total of 107 matching sources.
The discrepancy in the observed flux was less than 4.8 per cent, while the deboosted flux density was approximately 2.3 per cent. At the deepest part of the field of view, where the two observations overlap most, the differences were 0.4 and 2.9 per cent, respectively.

Employing the same matching range and radius, we also compared the flux of the SCUBA-2 SSA22 field with the deeper ALMA observations \citep[ADF22;][]{umehata2018}.
We compared the total flux contribution of the corresponding ALMA SMGs to the deboosted flux of the SCUBA-2 source and found that they were in agreement.
Given that ALMA SMGs are situated in the deepest part of our observed field of view, we believe that the contribution from faint sources below the detection threshold may be smaller.
It is important to note that there was an error in the flux conversion due to using a conversion factor from 1.1 mm to 850 $\upmu$m for the ALMA sources.
Additionally, the absolute flux calibration uncertainty is 15 per cent for S2CLS survey \citep{geach2017} and 10 per cent for ADF22 \citep{umehata2018}. In conclusion, we find that both the observed flux and deboosted flux are consistent with the submillimetre flux studied previously.

The final source catalogue consists of 390 submillimetre sources with a significance greater than 3.5$\sigma$, among which 92 sources reached a 5$\sigma$ level. The Table~\ref{tab:source_cata} presents the first 10 most significant detections, including information about source coordinates, measured and deboosted flux densities, signal-to-noise ratios, completeness, false detection rates, and positional offsets. The total uncertainty of deboosted flux combines the deboosting uncertainty, instrument noise, and confusion noise. For visual convenience, the values of FDR shown in the table are presented in logarithmic form. Caution should be exercised when using sources with high false detection risk and location uncertainty for follow-up observation and identification.

We plotted the distribution of deboosted sources in Fig.~\ref{fig:number_distribution}. The `deep region' corresponds to the area with $\sigma_\text{rms}$ $\leqslant$ 1 mJy in the science map, which is the overlapped region between new and archival observations. The dark grey bars in the plot represent the number distribution of sources in this region, which contains 171 sources. Additionally, the light grey bars stacked at the top display the distribution of 219 sources in the remaining area.

Among the sources, five have flux densities greater than 10 mJy, three of which fall between 10 and 11 mJy. The other two are the brightest sources in the SSA22 850 $\upmu$m map, with flux densities of 15.03 and 13.65 mJy, respectively. The former source is listed in the S2CLS catalogue presented by \citet{geach2017} and is located in a protocluster core with a redshift of $\sim$ 3.09 \citep{tamura2010, umehata2015} in the centre of the field. The latter source is found at the edge of the map and requires further investigation through multiwavelength studies to determine its nature, such as whether it is a local emitter, a lensing galaxy, a bright starburst galaxy, or a multiplicity.

\section{NUMBER COUNTS AND DISCUSSION}
\label{sec:counts_and_discussion}

\subsection{Number counts}
\label{sec:number_counts}
The surface number density of submillimetre galaxies serves as an effective test for galaxy formation models and provides valuable constraints on cosmological simulations. The submillimetre survey covers a larger area, then the map will have a smaller cosmic variance and the number counts will be closer to unbiased. A JCMT's ongoing large program, the SCUBA-2 Large eXtragalactic Survey \citep[S2LXS,][]{garratt2023}, is making significant progress in submillimetre survey area.

To construct the number counts for the SSA22 field, we followed a three-step process. First, we deboosted the flux densities of sources using the two-dimensional diagram of the simulation result (Section~\ref{sec:source_extraction}) to obtain the intrinsic flux density. 
Second, we calculated the surface density of each source by dividing one by the effective area and bin width, measured in units of deg$^{-2}$ mJy$^{-1}$ \citep{shim2020}. And we applied correction to each source based on its deboosted flux density, which involved weighting each source according to its authenticity and completeness:
\begin{equation}
    D_{i}=\frac{1}{{A}_e \times \text{d}S_i} \times \frac{1-\mathcal{F}_i}{\mathcal{C}_i},
    \label{eq:surface_density}
\end{equation}
where $A_e$ represents the effective area, and $\text{d}S_i$ denotes the width of the flux bin in which the deboosted flux density of each source is located. $\mathcal{F}_i$ and $\mathcal{C}_i$ stand for the false detection rate and completeness correction, respectively \citep{shim2020,zhang2022}. To obtain the raw number counts, we summed up the weighted surface number densities of all sources within each bin.

Finally, we adopted a technique from \citet{geach2017} to refine the raw number counts. Due to the presence of instrument noise, different injected flux densities could be measured as the same flux density, as explained in Section~\ref{sec:simulation_method}. Hence, we performed 1000 random samplings of the flux density for each source from its corresponding `intrinsic' (injected) flux density distribution generated by the Monte Carlo simulation (see Section~\ref{sec:flux_boosting}).
For each sampled flux density, the completeness correction values were computed based on the flux value and the local noise of the source. Conversely, the differential FDR corrections remained the same for all sampling fluxes of a particular source, as they depended solely on the observed S/N.
The surface number density within each flux density bin was calculated from the 1000 sampling processes. The mean of each flux bin was then considered the final differential number counts, with the standard deviation providing the uncertainty.

We also constructed cumulative number counts based on this `smoothing' process. This cumulative representation offers a valuable tool for quantifying the overall number density distribution of the 850 $\upmu$m sources and facilitates comparison with other submillimetre surveys or models.
The number counts, ranging from 2 to 16 mJy, were summarized in Table~\ref{tab:counts}. Here, S represents the bin centre, and S$^{\prime}$ represents the left edge of the flux bin. The cumulative counts described the number densities of sources with flux greater than S$^{\prime}$, with their uncertainties derived from the 1000 sampling processes.

\begin{figure*}
	\includegraphics[width=2\columnwidth]{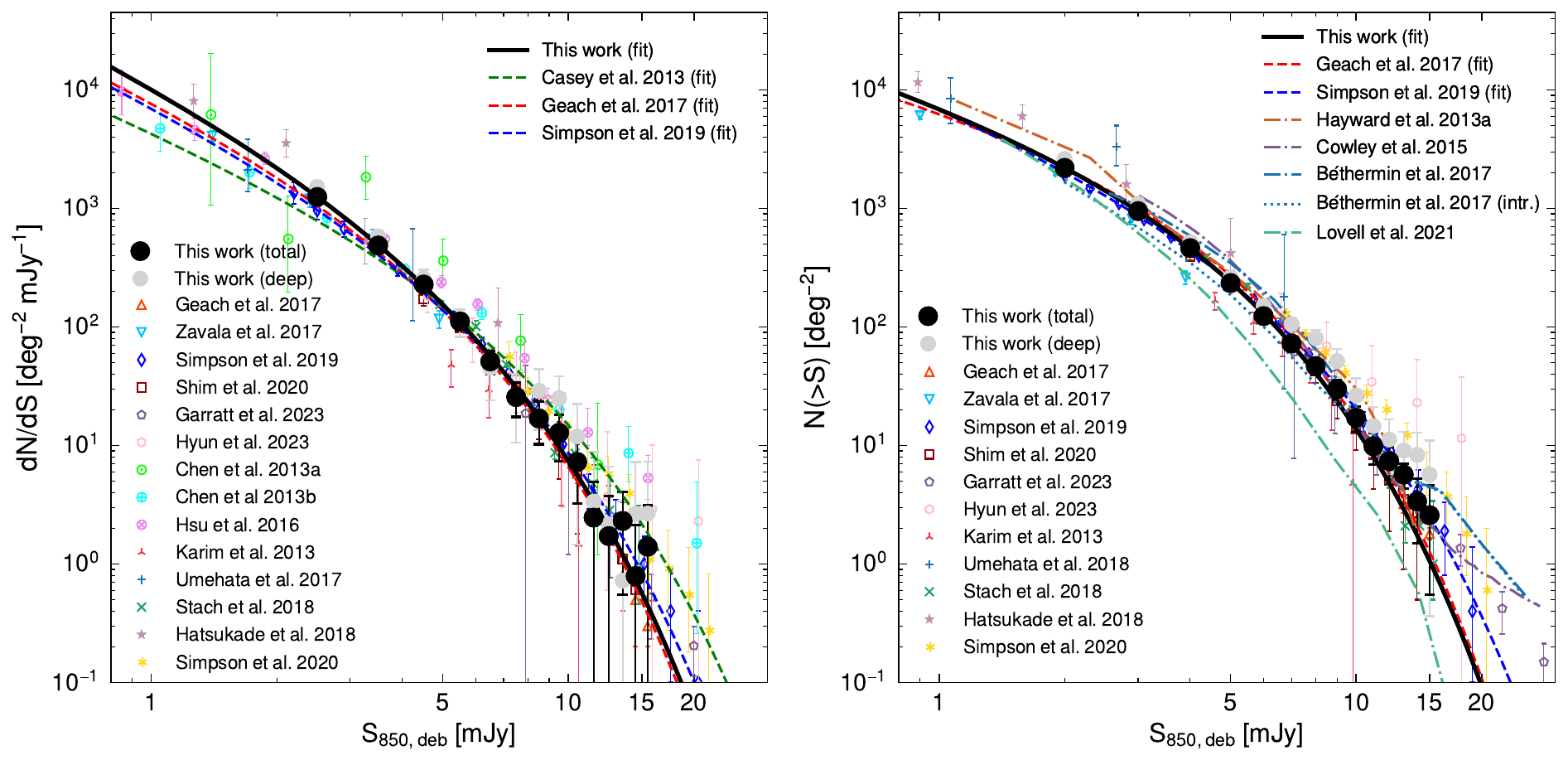}
    \caption{ 
    The left panel displays the differential number counts, while the right panel illustrates the cumulative counts.
    The black dots correspond to the number counts encompassing the entire effective area, accompanied by a solid black line representing the best-fitting Schechter curve. The grey dots depict findings from the central dense region with $\sigma_\text{rms}$ $\leqslant$ 1 mJy.
    Hollow polygons portray SCUBA-2 850 $\upmu$m number counts from single-dish surveys, while circles represent counts from fields influenced by the lensing effect. Branch markers indicate interferometric counts of ALMA at wavelengths of 870 $\upmu$m, 1.1 mm, and 1.2 mm, scaled to the 850 $\upmu$m band using a factor of $\nu^{3.8}$.
    Colorful dashed lines denote best-fitting counts from representative surveys conducted in the COSMOS field \citep{casey2013}, S2CLS \citep{geach2017}, and S2COSMOS \citep{simpson2019}.
    Through extensive comparison with results from other surveys, the number counts in the SSA22 field exhibit significant concordance with previously published counts.
    In the cumulative counts figure, we also present predictions of 850 $\upmu$m derived from various galaxy formation and evolution models, depicted as dot-dashed curves. These models encompass the semi-empirical model \citep{hayward2013a}, the semi-analytical model GALFORM \citep{cowley2015}, the empirical model SIDES \citep{bethermin2017}, and the cosmological hydrodynamical model SIMBA \citep{lovell2021}.
    Except for the model proposed by \citet{hayward2013a}, the remaining models account for the source blending effect introduced by the relatively large beam of 850 $\upmu$m submillimetre observations. The model by \citet{bethermin2017} constrains the SFR to values less than 1000 M$_\odot$ yr$^{-1}$ and achieves simulation outcomes that align well with observational data. We present both the source extraction number counts of their model and the intrinsic counts denoted by a blue dotted line.}
    \label{fig:counts}
\end{figure*}

The real number counts are illustrated in Fig.~\ref{fig:counts}, with black solid circles representing the counts for the entire effective area. The best-fitting line with the Schechter form for the differential number counts is indicated by the black solid line, featuring the following parameters: N$_0$ = 7785 $\pm$ 1095 deg$^{-2}$, S$_0$ = 2.64 $\pm$ 0.2 mJy beam$^{-1}$, and $\gamma$ = 1.65 $\pm$ 0.1.
As anticipated, our results closely align with the average number counts from \citet{geach2017}. This finding suggests that the number counts of the SSA22 field remain similar to those of blank fields, even in the presence of dense regions. While the Schechter model offers more physical insight \citep{geach2013}, we also provide the best-fitting results using the double-power law form:
\begin{equation}    
	\frac{\text{d}N}{\text{d}S}=\left(\frac{N_0}{S_0} \right)	\left[ \left( \frac{S}{S_0}\right)^{\upalpha}+\left( \frac{S}{S_0} \right)^{\upbeta} \right]^{-1}.
    \label{eq:double_powerlaw_form}
\end{equation}
Both fitting results are consistent with each other, despite their different functional forms. The parameters for both fittings are listed at the end of Table~\ref{tab:counts}.

\begin{table}
	\caption{
	The number counts for the SSA22 field and the best-fitting parameters of differential counts. The flux density S refers to the centre of the flux bin, while S$^{\prime}$ denotes the left edge of the flux bin. The column labelled d$N$/d$S$ represents the differential number counts, and $N$( > $S^{\prime}$) denotes the cumulative number counts.
    The uncertainties of the number counts arise from the standard deviations obtained from 1000 times of sampling. The last two rows display the best-fitting parameters for the differential counts. The first row presents the parameters for the Schechter form fitting, and the second row provides the parameters for the double power law form.}
	\label{tab:counts}

	\begin{tabular}{cccc}
	
	    \hline
	    S     & d$\textit{N}$/d$\textit{S}$ & S$^{\prime}$ & $\textit{N}$( > S$^{\prime}$) \\
	    (mJy) & (deg$^{-2}$ mJy$^{-1}$)       & (mJy)        & (deg$^{-2}$)                \\
	    
	    \hline
	    2.5  &  1249.0 $\pm$ 143.1 & 2.0  & 2198.8 $\pm$ 128.7 \\
	    3.5  &   487.0 $\pm$ 50.9  & 3.0  &  949.9 $\pm$ 46.6  \\
	    4.5  &   228.1 $\pm$ 28.5  & 4.0  &  462.9 $\pm$ 28.3  \\
	    5.5  &   111.2 $\pm$ 18.1  & 5.0  &  234.8 $\pm$ 18.3  \\
	    6.5  &    51.2 $\pm$ 11.3  & 6.0  &  123.6 $\pm$ 12.4  \\
	    7.5  &    25.6 $\pm$ 8.2   & 7.0  &   72.4 $\pm$ 8.7   \\
	    8.5  &    16.9 $\pm$ 6.7   & 8.0  &   46.8 $\pm$ 7.0   \\
	    9.5  &    12.8 $\pm$ 5.5   & 9.0  &   29.9 $\pm$ 5.7   \\
	    10.5 &     7.3 $\pm$ 4.0   & 10.0 &   17.1 $\pm$ 4.3   \\
	    11.5 &     2.5 $\pm$ 2.4   & 11.0 &    9.9 $\pm$ 3.0   \\
	    12.5 &     1.7 $\pm$ 2.0   & 12.0 &    7.4 $\pm$ 2.3   \\
	    13.5 &     2.3 $\pm$ 1.8   & 13.0 &    5.7 $\pm$ 1.3   \\
	    14.5 &     0.8 $\pm$ 1.3   & 14.0 &    3.4 $\pm$ 1.9   \\
	    15.5 &     1.4 $\pm$ 1.7   & 15.0 &    2.6 $\pm$ 2.0   \\
	    \hline
	    \hline
	    \multicolumn{4}{c}{\textit{best fitting parameters} } \\
	    N$_0$        &  S$_0$             &  $\gamma$ ($\alpha$) & $\beta$ \\
	    (deg$^{-2}$) &  (mJy beam$^{-1}$) &                      &         \\
	    \hline
	    7785 $\pm$ 1095 & 2.64 $\pm$ 0.2  & 1.65 $\pm$ 0.1  &              \\
	    642  $\pm$ 99  & 6.72 $\pm$ 0.30 & 2.62 $\pm$ 0.07 & 6.47 $\pm$ 0.53 \\
	    
	   \hline
	\end{tabular}

\end{table}

We observed an upturn in our number counts above 10 mJy, with the two brightest sources mentioned earlier (Section~\ref{sec:source_catalogue}) being the primary reason for the count points in the tail being higher than the prediction of the fitting curve. 
According to \citet{geach2017}, this upturn could be influenced by local sources or gravitational lensing galaxies.
In the deep region ($\sigma_\text{rms}$ $\leqslant$ 1 mJy; area $\sim$ 0.09 deg$^2$), represented by the light grey spots, we observed an apparent boost between 8 and 10 mJy (also see Fig.~\ref{fig:number_distribution}), aside from the excess caused by the known brightest source.

\subsection{Discussion}
\subsubsection{Comparison with other surveys}
In Fig.~\ref{fig:counts}, we compared our number counts with several published submillimetre surveys. Our differential number counts largely overlap with the S2CLS survey \citep{geach2017}, which covers an area of approximately 5 deg$^2$ and has a sensitivity of about 1.2 mJy. This indicates that it is reasonable to adopt their Schechter counts model for our Monte Carlo simulation. The S2LXS survey \citep{garratt2023}, which provides number counts of the rare bright source population in a uniformly scanned sky area of about 9 deg$^2$, also shows consistent results with ours within the flux density range we have presented. Moreover, the number counts of the SSA22 field are in close agreement with the source number density of the S2COSMOS \citep{simpson2019} and NEPSC2 \citep{shim2020}, both of which map a sky zone of about 2 deg$^2$. The S2COSMOS survey reaches a depth of approximately 0.5 mJy over a limited portion of the large survey field, while the EGS deep field map \citep{zavala2017} has a central instrumental noise of 0.2 mJy. And the sensitivity in the deepest region of the JWST-TDF survey \citep{hyun2023} reaches 0.8 mJy , similar to ours, and their survey estimations are essentially consistent with our counts, except for a slight excess at brighter end.

On the other hand, we also present the differential number counts of the lensing field A370 \citep{chen2013a}, which show values two or three times higher than our observations at flux densities $\gtrsim$ 2 mJy. 
\citet{chen2013b} and \citet{hsu2016} combined cluster lensing fields with blank fields, benefiting from the remarkable amplification of the gravitational lensing effect, extending significantly into the extreme faint flux regime of $\sim$ 0.1 mJy. Our statistical distribution and the extrapolation of the Schechter model at the faint end agree well with their consequences. 
For flux densities > 5 mJy, the lensing effect becomes evident, showing enhancements of ten to tens of times when the flux exceeds 10 mJy. This phenomenon is consistent with the conclusions reported by \citet{negrello2010}, \citet{vieira2010}, and \citet{aretxaga2011}, where a clear upturn is observed in the counts at the brighter flux range due to gravitational lensing amplification from foreground structures or cosmic variance.

Meanwhile, we presented and compared the ALMA 870 $\upmu$m, 1.1 mm, and 1.2 mm interferometric counts. To transform the counts to the 850 $\upmu$m band for visual comparison, we used the equation (7) from \citet{dunne_eales2001} and equation (4) from \citet{simpson2019}:
\begin{equation}    
	\frac{S_\text{850}}{S_\text{other}}=
	\left(\frac{\nu_\text{850}}{\nu_\text{other}}\right)^{\beta} \times
    \frac{B(\nu_\text{850}, T)}{B(\nu_\text{other}, T)},
    \label{eq:transform_diffwave}
\end{equation}
which relate the flux ratio at different (sub)millimetre bands to thermal dust emission and dust emissivity. For the dust emissivity $\beta$, we adopted a value of 1.8 \citep{plancklab2011, dudzeviciute2020}. Since the submillimetre regime is on the Rayleigh–Jeans side of the Planck function \citep{dunne_eales2001}, the second term in the equation~(\ref{eq:transform_diffwave}) is simply $\propto \nu^2$. Therefore, we used the scale conversion factor of $\nu^{3.8}$ to convert the interferometric results to the 850 $\upmu$m band for comparison. The expected S$_{850}$/S$_{1100}$ ratio was approximately 2.66, similar to conversion factors obtained from using the free spectral index parameter \citep{austermann2010} and assuming a simple empirical relation between the stellar formation rate and the gas-phase metal mass \citep{hayward2013a}. The multiplicative factor of $\sim$ 1.09 converted 870 $\upmu$m counts to corresponding 850 $\upmu$m flux bins, close to the value derived from the SED template assuming submillimetre galaxies located at z = 2 \citep{bethermin2017}.

The earliest report from \citet{karim2013} found that interferometric counts were generally consistent with results from low-resolution single-dish surveys, but falling below 850 $\upmu$m single-dish counts for flux regime less than 7 mJy and greater than 10 mJy. This suggested that most single-dish submillimetre sources were not caused by a mixing of unresolved SMGs, while bright sources were significantly affected by source blending.
\citet{stach2018} pointed out that interferometric observations were not affected by blending, and the counts dropped by 28 per cent compared to single-dish surveys. They and \citet{simpson2020} concluded that about half or more of single-dish sources with 850 $\upmu$m flux greater than 9--12 mJy were multiple sources. They also believed that about a third of these populations were physically related, which was consistent with the summary from galaxy formation models \citep[see Section~\ref{sec:compare_model};][]{hayward2013a, lovell2021}.

\citet{umehata2017} obtained the ALMA number counts of the SSA22 central region of 6 arcmin$^2$ (ADF22). We plotted their counts excluding the sources of the protocluster in Fig.~\ref{fig:counts} left panel. Based on this, \citet{umehata2018} extended observations to obtain a continuous ALMA interferometer map of 20 arcmin$^2$, whose showed several times more counts compared to the blank field. \citet{hatsukade2018} presented the 26 arcmin$^2$ ALMA blank field survey in the GOODS-S field, and their estimates were consistent with other single-dish surveys or interferometric counts.

Overall, our number counts were found to be in good agreement with previously published counts observed from submillimetre single-dish and interferometer surveys, as well as the lensing fields.

\subsubsection{Comparison with models}
\label{sec:compare_model}
The replication of high-redshift submillimetre galaxy counts has constituted a pivotal focus within galaxy formation models \citep{lovell2021}. In this section, we undertake a comparison between our cumulative counts and model predictions.

\citet{hayward2013a} employed a combination of a semi-empirical model, three-dimensional hydrodynamical simulations, and dust radiative transfer to predict SMGs counts. They also considered the contribution of individual subpopulations of SMGs, such as merger-induced starbursts, galaxy pairs, and isolated disc galaxies. 
Their simulated single-dish counts matched total cumulative counts at S$_\text{850}$ < 5 mJy and were in good agreement with our deep region counts ranging from 3 < S$_\text{850}$ < 12 mJy (refer to Fig.~\ref{fig:counts}). Despite not accounting for the lensing effect, their projections aligned with the bulk of observed counts in the high flux regime (> 10 mJy). 
\citet{hayward2013b} further considered and reevaluated the source blending due to a large beam size, finding that over half of blended sources had at least one spatially independent blending component, with a typical redshift separation of about 1.

\citet{lacey2016} introduced an enhanced iteration of GALFORM, a semi-analytical framework for galaxy formation and evolution. This updated model integrated a mildly top-heavy initial mass function (IMF), AGN feedback, a dust radiation model, and observational constraints to attain a more physically motivated representation. In the model, starbursts predominantly arose from disc instabilities. 
Utilizing the GALFORM model and employing the light-cone technique, \citet{cowley2015} simulated the 15 arcsec beam effect of JCMT, instrument noise, and matched filters to yield the `submillimetre map' at 850 $\upmu$m. They subsequently employed the `top-down' approach for source extraction. 
The counts derived from the simulated map, depicted in Fig.~\ref{fig:counts}, slightly exceeded the observed counts, manifesting an excess by a factor of $\sim$ 1.3--1.5 within the flux range of < 8 mJy. Model counts closely matched our counts at higher flux levels.

The empirical SIDES model \citep{bethermin2017} used updated version of the two star-formation modes \citep[i.e. main sequence star-forming and starburst galaxies; see][]{bethermin2012}, combined with observational constraints, realizing a 2 deg$^{2}$ extragalactic submillimetre survey simulation. The model imposed an upper limit of 1000 M$_\odot$ yr$^{-1}$ on the star formation rate (SFR). 
We plotted the simulated source extraction and intrinsic number counts with the SFR limit in Fig.~\ref{fig:counts}. Apart from extremely bright flux densities (> 15 mJy), the source extraction counts remain consistent with observed counts across the flux density range. Between S$_\text{850}$ = 5--10 mJy, the model overpredicts cumulative counts by around 50--60 per cent. 
The intrinsic counts curve is lower than our observed counts by $\sim$ 30--50 per cent for S$_\text{850}$ < 10 mJy, aligning with interferometric observation results \citep{simpson2015,stach2018}. At the brighter end (> 10 mJy), intrinsic counts are in agreement with single-dish observations.

\citet{lovell2021} harnessed the cosmological hydrodynamic simulation SIMBA and dust radiation transfer to generate 850 $\upmu$m emission. To consider the source blending effect, they projected the radiation from light-cone onto a 0.5 deg$^2$ sky plane, which was then convolved with the SCUBA-2 PSF to generate a simulated single-dish `observed map'. 
Their predicted cumulative counts lie a factor of about 1.3--2 below observed counts at S$_\text{850}$ = 2--5 mJy, and 2--3 times lower than our measurements at S$_\text{850}$ = 5--10 mJy.

\subsubsection{SMGs overdensity ?}
\label{sec:counts_excess}

In this section, we aim to analyse the variation in our field compared to the blank field and quantify the excess of submillimetre sources in the SSA22 field.
Before proceeding, we will first estimate the cosmic variance of the SSA22 field. Utilizing the configuration proposed by \citet{moster2011}, we adopted a mean redshift of z = 3.09 and a redshift bin width of $\Delta$z = 1. The cosmic variances for typical submillimetre galaxies (with logM$_{\star}$/M$_{\odot}$ $\sim$ 10.5) in the total map and the deep region are approximately 19 per cent and 26 per cent, respectively. Consequently, even within the central region, the cosmic variance remains insignificant.
When considering solely the extremely narrow redshift slice at redshift 3.09 ($\Delta$z = 0.1), cosmological variance becomes somewhat significant. 
However, due to the broad range of redshifts (z $\sim$ 1 - 6) observable in the submillimetre waveband, and our extensive coverage, the overall detected cosmic volume is vast, diminishing the impact of cosmic variance \citep{somerville2004, driver2010}. 
Based on these results, it appears that the large-scale structure of SSA22 has a limited impact on the counts, suggesting that the fluctuations in the counts are likely primarily due to Poisson randomness.

\citet{simpson2019} suggests that cosmic variance does not strongly affect the 850 $\upmu$m population on scales of $\sim$ 0.5--3 deg$^{2}$. Therefore, it is reasonable to regard the average number counts of S2CLS survey ($\sim$ 5 deg$^{2}$ ) as a representative of blank field, which we use to estimate the excess of SSA22 850 $\upmu$m counts.
To conduct the comparison, we divided the astronomical map into two regions: the `deep region' with $\sigma_\text{rms}$ $\leqslant$ 1 mJy, located at the centre of the map and containing the protocluster core, and the `shallow region' (1 mJy < $\sigma_\text{rms}$ < 2 mJy) covering the rest of the area. These two regions occupy $\sim$ 0.09 and 0.25 deg$^{2}$, respectively. Following the approach used by \citet{geach2017}, we used the equation $\Delta$ = (N — N$_{\text{S2CLS}}$) / N$_{\text{S2CLS}}$ to assess the elevation in source density, where N and N$_{\text{S2CLS}}$ represented the cumulative number counts of the SSA22 field and the S2CLS survey, respectively. The value of $\Delta$ represents the cosmic variance of number counts of each region compared to the blank field.

\begin{figure}
	\includegraphics[width=\columnwidth]{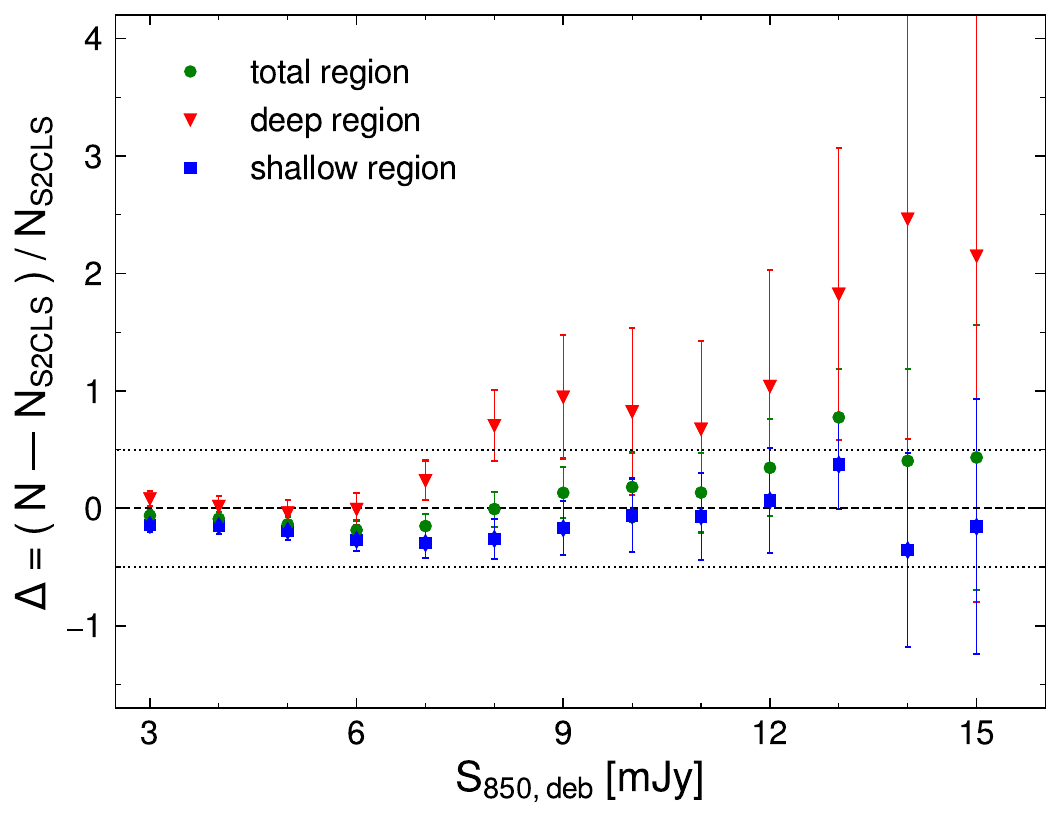}
    \caption{ 
    The cosmic variance of 850 $\upmu$m number counts in the SSA22 field compared to the blank field. The red inverted triangles, blue squares, and green circles represent the deviation of cumulative number counts in the deep region, shallow region, and total region, respectively, relative to the mean density of the S2CLS survey. The dotted lines correspond to 50 per cent of the S2CLS average source density.
    Compared to the blank field counts, the deep region exhibits an obvious excess, while the shallow region shows good agreement with the S2CLS survey. The total number counts lie somewhere in between, and it is reasonable to attribute this discrepancy to the concentration of sources in the deep region at the field centre.}

    \label{fig:delta_cumu_counts}

\end{figure}

In Fig.~\ref{fig:delta_cumu_counts}, the dotted lines represent 50 per cent of the S2CLS average surface number density. 
The counts in the shallow region are well consistent with the 50 per cent ranges of the S2CLS source density, indicating that the shallow region is comparable to a blank field. 
The total number counts in the SSA22 field are also good consistent with the blank fields in the flux range of < 10 mJy.
In the deep region, for fluxes greater than 10 mJy, the counts seem to exceed those of the blank field by a considerable margin. However, given the limited number of extremely bright sources in this range, the impact of randomness is relatively significant. Additionally, as discussed in Section~\ref{sec:number_counts}, this upturn may be influenced by local or lensing galaxies. Furthermore, the impact of cosmic variance in the deep region is relatively limited. Therefore, we believe that the apparent excess in this flux regime may not necessarily indicate a significant influence of large-scale structure on the counts and this variation is more likely due to Poisson noise \citep{geach2017}.
In the remaining part of the deep region counts, there is only a slight excess at around 9 mJy, we conducted a simple comparison of the source number density with the cumulative curves presented in \citet{geach2017} and \citet{simpson2019}. We calculated the source number density N$_\text{SSA22}$ / N$_\text{blank}$ for a specific flux cut, where N$_\text{SSA22}$ represents the source number density in our field, and N$_\text{blank}$ represents that from the blank field. 
At S$_{850}$ $\geqslant$ 4 mJy, the ratios for the total and deep regions are 0.97 $\pm$ 0.06 and 1.08 $\pm$ 0.10, respectively. At S$_{850}$ $\geqslant$ 9 mJy, these ratios become 1.03 $\pm$ 0.20 and 1.77 $\pm$ 0.48.

Following the approach of \citet{battaia2018} and \citet{zhang2022}, we quantified the  excess of the 850 $\upmu$m number counts by comparing them to the blank fields represented by S2CLS \citep{geach2017} and S2COSMOS \citep{simpson2019}. 
We refitted our differential counts and derived the best-fitting parameter N$_{0}$ using their parameters S$_0$ and $\gamma$ of the Schechter counts model. When fixing S$_0$ = 2.5 and $\gamma$ = 1.5 \citep{geach2017}, we found N$_\text{0, refit}$ / N$_\text{0, S2CLS}$ ratios of approximately 1.33 and 1.04 for the deep and shallow counts, respectively. Fixing S$_0$ = 3.0 and $\gamma$ = 1.6 \citep{simpson2019}, the N$_\text{0, refit}$ / N$_\text{0, S2COSMOS}$ ratios were approximately 1.38 and 1.08 for the deep and shallow counts.
For direct comparison with known overdensity fields, we selected smaller central areas with $\sigma_\text{rms}$ $\lesssim$ 0.84 and 0.95 mJy ($\sim$ 0.036 deg$^2$ and 0.074 deg$^2$) in the SSA22 field, which roughly corresponded to the sizes of individual fields presented by \citet{battaia2018} and \citet{zhang2022}. In these smaller regions, the density excesses in the counts compared to blank fields were about 48 per cent and 25 per cent, respectively, much lower than the estimations of 4 times and 2 times the number density of the blank field mentioned in the literature.
In brief, the counts in the deep region are 1.35 times those of the blank field, while the total counts are only 1.16 times.

Overall, within our submillimetre waveband detection volume, cosmic variance is not substantial. 
The 850 $\upmu$m number counts, derived from a total detection coverage of approximately 0.3 deg$^2$, conform to the blank field. 
In the deep region, the number counts for flux density < 8 mJy are in agreement with the blank field. 
Due to the lower number of sources at the brighter end of the flux density and the insignificant cosmic variance, we believe that the excess at the brighter end is mainly influenced by Poisson noise. 
The quantified excess of the deep region counts over the blank field is 1.35 times, which is not statistically significant.
In the SSA22 field, the detection of SCUBA-2 sources at 850 $\upmu$m does not exhibit signs of overdensity.

\section{Summary}
In this work, we present the deepest 850 $\upmu$m mosaic map of the SSA22 field to date. The map is a combination of JCMT/SCUBA-2 new observations and archival data spanning from 2012 September to 2015 June, totaling 91 hours of observations. The sensitivity of the map is limited to less than 2 mJy beam$^{-1}$, resulting in a total effective area of about 0.34 deg$^2$. The deep region, with an $\sigma_\text{rms}$ $\leqslant$ 1 mJy, covers approximately 0.09 deg$^2$, while the central deepest coverage reaches an rms of $\sim$ 0.79 mJy beam$^{-1}$, which is about 1.5 times deeper than the previous S2CLS survey. 
We estimate the confusion noise to be 0.43 mJy beam$^{-1}$.

We process the raw data and adopt an iterative `top-down' method to extract the potential submillimetre sources. The extraction threshold is set to 3.5$\sigma$, meaning that 17 per cent of sources are false detections. The final source catalogue contains 390 sources, with 92 sources reaching 5$\sigma$. This SSA22 850 $\upmu$m source catalogue is publicly available. We perform a Monte Carlo simulation 5000 times using the jackknife true noise map to estimate the effects of flux boosting, positional uncertainty, completeness, and false detection rate. Our flux boosting and positional uncertainty are larger than S2CLS due to differences in mapping uniformity, survey area, and cosmic variance. The mean value of boosting factors for all sources is about 1.4, and the average positional uncertainty is about 2.5 arcsec. The average completeness at 50 and 80 per cent levels is 3.7 and 5.1 mJy, respectively. At the 4$\sigma$ level, the cumulative false detection rate is about 5 per cent, which drops to approximately 1 per cent when the source significance reaches 4.5$\sigma$.

We construct the SSA22 850 $\upmu$m number counts, showing good agreement with some large-area SCUBA-2 surveys, such as S2CLS, S2COSMOS, and NEPSC2 \citep{geach2017, simpson2019, shim2020}, within the deboosted flux range of 3--15 mJy. Additionally, the counts are broadly consistent with published 850 $\upmu$m measurements from single-dish and interferometer surveys, and lensing field, as well as predictions from galaxy formation models. 
We observe a clear upturn in the counts in the brighter flux regime. 

Compared to the blank fields, we measure that the total effective area counts are 1.16 times higher, and the counts in the deep region show an improvement of about 1.35 times.
In the flux density range of < 8 mJy, the deep region closely resembles the blank field. 
Due to the limited number of sources at the brighter end and the insignificant impact of cosmic variance from the large-scale structure, we attribute the excess of deep region counts in the brighter flux regime mainly to Poisson noise. 
In general, the total number counts align with the blank field, and the excess of the deep region counts relative to the blank field is not significant.
Furthermore, we find that the number counts do not indicate overdensity.

Given the abundant multiwavelength data sets available for the SSA22 field, including X-ray, optical, infrared, and radio bands, we plan to conduct research on the physical properties of the submillimetre sources in the SSA22 field, and to explore the evolution process of high-redshift galaxies and the structure of the early universe.

\section*{Acknowledgements}
We express our sincere gratitude to the anonymous referee for providing a constructive report that has enhanced the quality of this paper. Y.A. acknowledges the support from the National Key R\&D Program of China (2023YFA1608204), the National Natural Science Foundation of China (NSFC grants 12173089, 11933011), the Natural Science Foundation of Jiangsu Province (BK20211401), and the “Light of West China” Program (No. xbzg-zdsys-202212).
The James Clerk Maxwell Telescope is operated by the East Asian Observatory on behalf of The National Astronomical Observatory of Japan; Academia Sinica Institute of Astronomy and Astrophysics; the Korea Astronomy and Space Science Institute; the National Astronomical Research Institute of Thailand; Center for Astronomical Mega-Science (as well as the National Key R\&D Program of China with No. 2017YFA0402700). Additional funding support is provided by the Science and Technology Facilities Council of the United Kingdom and participating universities and organizations in the United Kingdom and Canada. Additional funds for the construction of SCUBA-2 were provided by the Canada Foundation for Innovation. Y.Z. acknowledges ﬁnancial supports from the Agencia Estatal de Investigación del Ministerio de Ciencia e Innovación (AEIMCINN) under grant (La evolución de los cíumulos de galaxias desde el amanecer hasta el mediodía cósmico) with reference (PID2019105776GB-I00/DOI:10.13039/501100011033) and the China Scholarship Council (202206340048).

\section*{Data Availability}
All the raw data are available at the the Canadian Astronomy Data Centre (\href{https://www.cadc-ccda.hia-iha.nrc-cnrc.gc.ca/en/jcmt/}{https://www.cadc-ccda.hia-iha.nrc-cnrc.gc.ca/en/jcmt/}) under program IDs M15AI91 and MJLSC02.
The data underlying this article are available from the corresponding author upon reasonable request.

The full 850 $\upmu$m source catalogue of 3.5$\sigma$ is available in the online journal.



\bibliographystyle{mnras}
\bibliography{number_counts}





\appendix



\bsp	
\label{lastpage}
\end{document}